# Sleeping beauties and temporal evolution of the coronavirus literature

Milad Haghani[1,2*], Pegah Varamini[3]

[1]School of Civil and Environmental Engineering, The University of New South Wales, Australia
[2]Institute of Transport and Logistics Studies, The University of Sydney, Australia
[3]Faculty of Medicine and Health, The University of Sydney, Australia

[*]Corresponding author: milad.haghani@sydney.edu.au

Temporal evolution of the coronavirus literature over the last thirty years (N=43,769) is analysed along with its subdomain of SARS-CoV-2 articles (N=27,460) and the subdomain of reviews and meta-analytic studies (N=1,027). (i) The analyses on the subset of SARS-CoV-2 literature identified studies published prior to 2020 that have now proven highly instrumental in the development of various clusters of publications linked to SARS-CoV-2. In particular, the so-called "*sleeping beauties*" of the coronavirus literature with an awakening in 2020 were identified, i.e., previously published studies of this literature that had remained relatively unnoticed for several years but gained sudden traction in 2020 in the wake of the SARS-CoV-2 outbreak. (ii) The subset of 2020 SARS-CoV-2 articles is bibliographically distant from the rest of this literature published prior to 2020. Individual articles of the SARS-CoV-2 segment with a *bridging* role between the two bodies of articles (i.e., before and after 2020) are identifiable. (iii) Furthermore, the degree of bibliographic coupling within the 2020 SARS-CoV-2 cluster is much poorer compared to the cluster of articles published prior to 2020. This could, in part, be explained by the higher diversity of topics that are studied in relation to SARS-CoV-2 compared to the literature of coronaviruses published prior to the SARS-CoV-2 disease. This work demonstrates how scholarly efforts undertaken during peace time or prior to a disease outbreak could suddenly play a critical role in prevention and mitigation of health disasters caused by new diseases.

(manuscript submitted to *Scientometrics*)

## 1. Introduction

The influx of scholarly publication that has ensued from the outbreak of the 2019 Novel Coronavirus (2019-nCoV) has been labelled as unprecedented and marks a unique phenomenon in the history of scientific publications (Haghani and Bliemer, 2020; Homolak et al., 2020; London and Kimmelman, 2020). While the scientific literature of coronaviruses has displayed a spike in the number of publications following the onset of each major outbreak, including that of the first Severe Acute Respiratory Syndrome (SARS) and Middle Eastern Respiratory Syndrome (MERS) diseases, the surge in the number of publications related to the recent outbreak of SARS-CoV-2 has made previous spikes pale into insignificance (Chiu et al., 2004; Di Girolamo and Reynders, 2020; Kostoff and Morse, 2011; Wang et al., 2016) (See Figure 1 which visualises this trend). In fact, while scientific studies on coronaviruses have been being undertaken for more than 50 years, according to early publications documented since late 1960s (Almeida et al., 1968; Haghani et al., 2020), the size of this literature nearly tripled within approximately eight months since the onset of the SARS-CoV-2 outbreak.

It is a rather commonly observed phenomenon that large-scale public health emergencies and major disease outbreaks mobilise scientific efforts which could lead to sudden surges of scientific publications on the respective disease (Johansson et al., 2018; Sweileh, 2017; Zhang et al., 2020c). However, the magnitude of the bursts in publications associated with SARS-CoV-2 and the broad range of sources across which these studies have been disseminated remains, arguably, unmatched. In other words, while academics often respond quickly to public health emergencies, the response to the SARS-CoV-2 outbreak is like no other, at least in terms of its magnitude (Fazeli-Varzaneh et al., 2021; Sa'ed and Al-Jabi, 2020; Torres-Salinas et al., 2020).

This tremendous scientific effort, however, has also posed challenges to academics on multiple fronts. In addition to exerting exceeding pressure on the review processes and on editorial boards of the journals that are deemed significant outlets for these publications, the sudden growth of the literature has also made it challenging for individual researchers to keep pace with the growing new developments. An article published by *Science* in May 2020 highlighted this issue in these words: "scientists are drowning in COVID-19 papers" (p.1) while referring to this global scientific effort in the dissemination of outcomes related to COVID-19 as one of "the biggest explosions of scientific literature ever" (p.1) (Brainard, 2020). This has, for example, prompted researchers of other fields to also weigh in and respond to these challenges by making efforts parallel to those of medical scientists in order to create data-mining tools and search tools as well as automated research synthesis and literature mining tools powered by Artificial Intelligence that could help researchers stay abreast of the new developments in this literature (Brainard, 2020; Hope et al., 2020).

The abovementioned issues indicate that, in facing this novel phenomenon in scientific publications, the role of research synthesis could become more important than ever. In fact, as we will show in the following sections, the current work documents that out of all review and meta analytical publications that have thus far been published in the coronaviruses literature, nearly 88% have been prompted by the SARS-CoV-2 outbreak. This shows how, alongside the rapid and dynamic growth of this literature, parallel efforts needed to be undertaken to synthesise the new developments and to make the findings more readily accessible to researchers in this area.

The objectives of the current study are to (i) identify cohorts and streams of thematically similar publications based on measures such as the similarity of their references or the frequency of them being jointly cited, (ii) to identify individual studies that are most influential within each stream of studies, (iii) to provide insight into the temporal evolution of the coronavirus literature, and (iv) to identify studies of this literature published prior to 2020 that are currently proving instrumental in informing the abundance of articles emerging in the wake of the SARS-CoV-2 pandemic, particularly those that had remained relatively unnoticed and are now gaining sudden traction as a result of the SARS-CoV-2 outbreak (i.e., the so-called *sleeping beauties* of the coronavirus literature). Limited supplementary analyses are also conducted on the subset of review and meta-analytic papers of the coronavirus literature. The outcomes of these analyses will help researchers of this field better understand the structural composition of this literature, as well as its major stream of research activities and fundamental studies within each stream. It is expected that the study contributes to informing future synthesis of research in this field, such as scoping and narrative review articles, a cohort of studies that is becoming exceedingly important in this field.

## 2. Methods and data acquisition strategy

The underlying datasets for the analyses of this work have been sourced from the Web of Science (WoS) Core Collection. Three separate datasets were obtained, one representing the general literature related to coronaviruses (dataset (I)), one representing the literature related to SARS-CoV-2 (dataset (II)), and one representing the subset of review and meta-analytic articles of the coronavirus literature (dataset (III)). Term-based search strategies were devised to retrieve each dataset from the WoS. The search query strings associated with datasets (I), (II) and (III) are detailed below respectively as search query (I), (II) and (III).

**Search query (I)**: *"coronavirus*" OR "corona virus" OR "Covid-19" OR "covid19" OR "2019-ncov" OR "Severe acute respiratory syndrome" OR "Middle east respiratory syndrome" OR "SARS disease" OR "SARS virus" OR "MERS disease" OR "MERS virus"*

**Search query (II)**: *"covid-19" OR "covid19" OR "coronavirus disease 2019" OR "2019-nCov" OR "Novel Coronavirus" OR "Novel Corona virus" OR "SARS-Cov-2"*

**Search query (III)**: *TS=("coronavirus*" OR "corona virus" OR "Covid-19" OR "covid19" OR "2019-ncov" OR "Severe acute respiratory syndrome" OR "Middle east respiratory syndrome" OR "SARS disease" OR "SARS virus" OR "MERS disease" OR "MERS virus")* **AND** *(TI=("literature review" OR "literature survey" OR "literature mapping" OR "A review" OR "scoping review" OR "systematic review" OR "systematised\* review" OR "systematic survey of the literature" OR "comprehensive review" OR "critical review" OR "mapping review" OR "mixed methods review" OR "evidence synthesis" OR "synthesis of evidence" OR "rapid review" OR "state-of-the-art review" OR "review of the state of the art" OR "qualitative review" OR "quantitative review" OR "umbrella review" OR "meta analysis" OR "meta-analysis" OR "meta analytic" OR "meta synthesis" OR "meta-synthesis" OR "meta evidence") OR AK=("literature review" OR "literature survey" OR "systematic review" OR "scoping review" OR "critical review" OR "literature analysis" OR "literature mapping" OR "survey of literature" OR "meta-analysis" OR "meta research"))*

Each search targeted the title, abstract, author keywords and Keyword Plus of the indexed publications, with no restriction on the time span of the research, with the exception of dataset (II) for which the time span was limited to 2019-2021. The research for datasets (I) and (II) was conducted on 21 and 22 August 2020 which returned N=43,796 items and N=27,460 items respectively. These numbers could, therefore, be regarded as estimates of the size of the coronavirus and SARS-CoV-2 literature at that particular point in time as recorded by the WoS. The quotation marks embodying the search terms ensure that the term combinations are considered verbatim in the search. The asterisk sign that accompanies the term "coronavirus" allows for the plural version (and other possible variations) of the term to also be considered. While search query (I) and search query (II) were both devised for the Basic Search section of the WoS, search query (III) had features that made it necessary to utilise the Advanced Search section of the WoS. In this string, "TS" indicates Topic (i.e., title, abstract, author keywords and Keyword Plus), "TI" means Title, and "AK" means Author Keywords, as the domains of search. The reason for limiting the domains of search for terms associated with review articles to the titles and author keywords is merely to keep false positives (non-review papers) to a minimum. The logic underlying the formulation of this search strategy in this fashion was that it is unlikely that a review/meta-analysis paper on coronavirus makes no mention of the terms that characterise review papers in either the title or the keywords list (and do so only in the abstract). Therefore, in order to minimise the likelihood of false positives in the data, the domain of the search was limited to author keywords and titles. The first and last one hundred items identified by this search were manually screened to ensure that the search is not producing false positives, i.e., non-review papers. And pre-processing confirmed that the data has presents this intended quality. This search strategy returned N=1,027 items on 4 September 2020. Full details of the documents associated with each dataset were exported from the WoS in the form of text files and in bundles of 500 items. The data includes citation count and bibliographic information of individual documents (i.e. authors, document title, year, source title, volume, issue, page, citation count and affiliations) as well as their abstract, keywords and list of references.

Apart from the general statistics obtained from the data, the main focus of the analyses is on the bibliographic coupling and co-citation relations of documents (as discussed in subsequent sections). For the bibliographic coupling analyses, VOSviewer1.6.15.0 (Van Eck and Waltman, 2010) and for the co-citation analyses, CiteSpace5.7.R1 (Chen, 2006; Chen et al., 2012) was used. The study also uses the outcome of the document co-citation analysis on the SARS-CoV-2 subset to determine articles of the coronavirus literature—published prior to 2020—that have suddenly been found instrumental in the development of the SARS-CoV-2 literature and have therefore had a sudden surge in citation, i.e., the *sleeping beauties* (Du and Wu, 2018; Van Raan, 2017) of the literature. The identification of sleeping beauties was assisted by a document co-citation analysis on the subset of SARS-CoV-2 papers. The unique structure of the coronavirus literature (i.e., the concentration of the SARS-CoV-2 literature within a single year) allowed us to conduct a document co-citation analysis on the SARS-CoV-2 subset (i.e., the 2020 segment of the coronavirus literature) and determine articles published prior to 2020 that are frequently co-cited with SARS-CoV-2 papers and by SARS-CoV-2 papers. Such articles were taken as candidates for sleeping beauties. Their citation patterns were then analysed individually. The analyses on sleeping beauties is not claimed to have aimed for a comprehensive list of sleeping beauties of the coronavirus literature. The focus is merely on sleeping beauties that have had an awakening in 2020 following the SARS-CoV-2 outbreak. Possible sleeping beauties with awakening prior to 2020 (if any exists at all) will remain undetected by this analysis.

Our assumption in approaching this question is that the exceptional surge of publication on coronavirus-related topics in 2020 might have acted as awakening role for certain studies of this literature published prior to 2020. Therefore, by conducting a document co-citation analysis of the 2020 SARS-CoV-2 articles, we seek to identify articles published prior to 2020 that have been heavily cited by 2020 SARS-CoV-2 articles and alongside other 2020 SARS-CoV-2 items, thereby identifying potential sleeping beauties of this field with an awakening in 2020 (following the SARS-CoV-2 outbreak).

## 3. Analyses and results

*3.1. General statistics of the coronavirus literature*

The size of the coronavirus literature as indexed by the WoS and as recorded on 21 August 2020 is 43,796 documents. Of these, 56% are Article, 16% Editorial Material, 13% Early Access, and 13% Letter, among other document types (including 7 Retracted articles, which are all 2020 items related to SARS-CoV-2). These publications have been indexed across a broad range of categories based on the journals in which they have been published. The top six categories include *Virology* (12%), *Medicine General Internal* (11%), *Infectious Diseases* (8%), *Public Environmental Occupational Health* (6%), *Immunology* (6%) and *Veterinary Sciences* (5%). Note that the categories could be overlapping as a journal could be indexed in more than one category. The oldest article of this literature, as indexed by the WoS, dates back to 1968 (Almeida et al., 1968). The distribution of the number of publications in this literature over the years, since 1968, has been visualised in Figure 1. Nearly 35% of this literature has accumulated during more than 50 years from 1968 to 2019, while the rest, 65%, has emerged within the first eight months of 2020.

Among the sources of coronavirus publications, *Journal of Virology* with 1,191 recorded items is the main source of publications. That is followed by *British Medical Journal* (798 items), *Journal of Medical Virology* (572 items), *Virology* (528 items), *Advances in Experimental Medicine and Biology* (378 items), and *Emerging Infectious Diseases* (374 items). According to this record, there are 443 sources/journals with at least 20 publications on this topic. Authors affiliated with the organisations based in the United States of America have been authors or co-authors of nearly 12,800 articles on this topic (making up nearly 20% of all publications). This is followed by authors affiliated with organisations based in People Republic of China (7,602 items), England (3,901 items), Italy (3,690 items), Canada (2,252 items) and Germany (2,192 items). There are 52 countries whose scholars have been collectively listed in at least 100 publications on this topic. The list of organisations whose scholars have been most active in publishing articles of this literature include *The University of Hong Kong* (840 items), *The Huazhong University of Science and Technology* (589 items), *The Chinese Academy of Sciences* (585 items), *The University of Toronto* (524 items), *The Chinese University of Hong Kong* (512 items) and *Harvard Medical School* (487 items). There are 141 organisations whose authors have collectively published at least 100 documents on this topic.

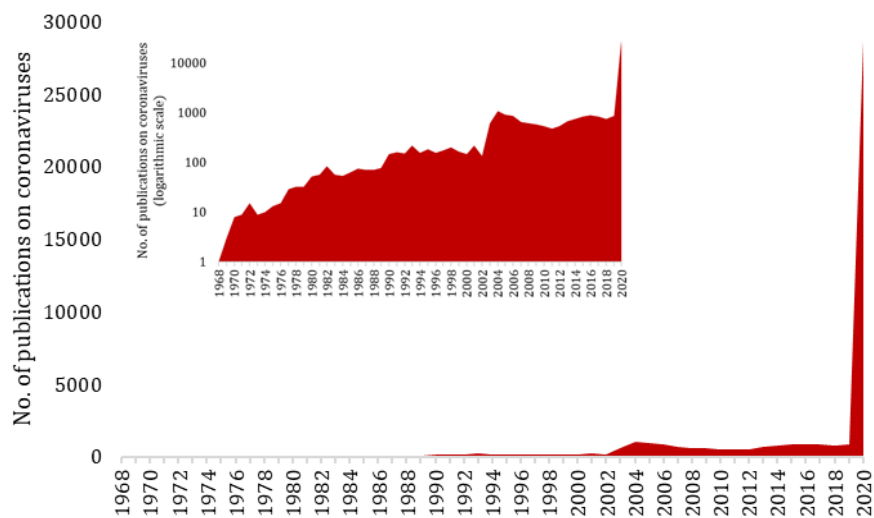

**Figure 1** The distribution of the number of publications related to coronaviruses over time since 1968. The data for 2020 only reflects the first eight months of the year. The figure embedded within the main graph visualises the same data on a logarithmic scale.

*3.2. Bibliographic coupling of coronavirus publications*

Thematic similarity of coronavirus publications could be explored through the number of references that they share, i.e., their bibliographic coupling. Two articles are deemed bibliographically coupled when they have a common reference, while the higher the number of such shared references, the stronger the bibliographic coupling relationship is. A map of bibliographic coupling identifies clusters of the articles that are related to one another based on this metric. Figure 2 visualises such map for documents of the coronavirus literature. In this map, each node represents a document in this literature, and the size of the node is proportional to the total number of global citations to that document. Documents with strongest bibliographic coupling relation are visualised at spatial proximity to one another. Also, links between pairs of nodes as well as the thickness of the links are indications of bibliographic coupling between pairs of articles. The map of bibliographically coupled coronavirus documents identifies three major clusters of articles, shown green, grey and purple. A hybrid version of this map that colour-codes the nodes based on their year of publication has also been shown at the top right corner of the main map. As can be seen from this hybrid map, the majority of documents in this literature are 2020 publications. The cluster on the right side of the map (shown green) consists mostly of a mixture of studies published prior to 2020. This cluster is bibliographically distant from the two other clusters that consist mostly of 2020 documents (shown purple and grey). However, a group of 2020 studies appear as a *bridge* between the two bodies of the coronavirus literature, linking the cohort of 2020 publications to those published earlier, meaning that they share references with both groups. Some of these bridging articles (or in other words, articles with *betweenness centrality* role (Barthelemy, 2004)) include those of Walls et al. (2020), Ou et al. (2020), Zimmermann and Curtis (2020), Wang et al. (2020c), Ashour et al. (2020), Fung et al. (2020), Ahn et al. (2020), Wu et al. (2020a), Shang et al. (2020). In fact, the titles of two of these articles— Ashour et al. (2020) ("Insights into the Recent 2019 Novel Coronavirus (SARS-CoV-2) in Light of Past Human Coronavirus Outbreaks") and Fung et al. (2020) ("A tug-of-war between severe acute respiratory syndrome coronavirus 2 and host antiviral defence: lessons from other pathogenic viruses")—are themselves indicative of their intended bridging nature. The most cited document of the 2020 clusters (i.e., purple and grey clusters, collectively) is the early study of Guan et al. (2020) first appearing on 28 February, 2020 ("Clinical Characteristics of Coronavirus Disease 2019 in China"). Other highly cited items include articles of Lai et al. (2020), Chen et al. (2020c), Young et al. (2020), Zhang et al. (2020a), Alhazzani et al. (2020). In the green cluster, the most prominent documents based on the metric of total (global) citations are Peiris et al. (2003c) and Lai and Cavanagh (1997). Note that the global citation count does not differentiate between citations that these articles have received from other coronavirus papers and those received from articles outside that field. Rather, it reflects a total count that includes both.

A striking observation based on the map of bibliographic coupling is the clearly poorer connectivity of the 2020 cluster(s) of articles compared to the degree of connectivity between items of this literature that have been published prior to 2020. The 2020 cluster is not only bibliographically distant from the rest of the body of this literature (with the exception of some individual 2020 articles that show certain level of betweenness), it is also highly dispersed within itself while the bibliographic connectivity between its individual articles is relatively poor. This could, in fact, be a reflection of the abundance of early studies on SARS-CoV-2 that have been undertaken almost concurrently and have been unloaded on this literature within the course of only few month, resulting in a large set of documents that do not show much bibliographic similarity compared to their counterpart articles published prior to 2020. It could as well, and to some degrees, be a reflection of the broader diversity of the topics investigated by 2020 cohort of the coronavirus studies compared to the literature that existed before this date. The nature of the journals across which the 2020 articles have been disseminated indicates a much greater range of topical diversity compared to the studies published prior to 2020 which could as well as be a sign of broader diversity of their underlying topics and, by extension, a plausible explanation for the poorer bibliographic connectivity between them.

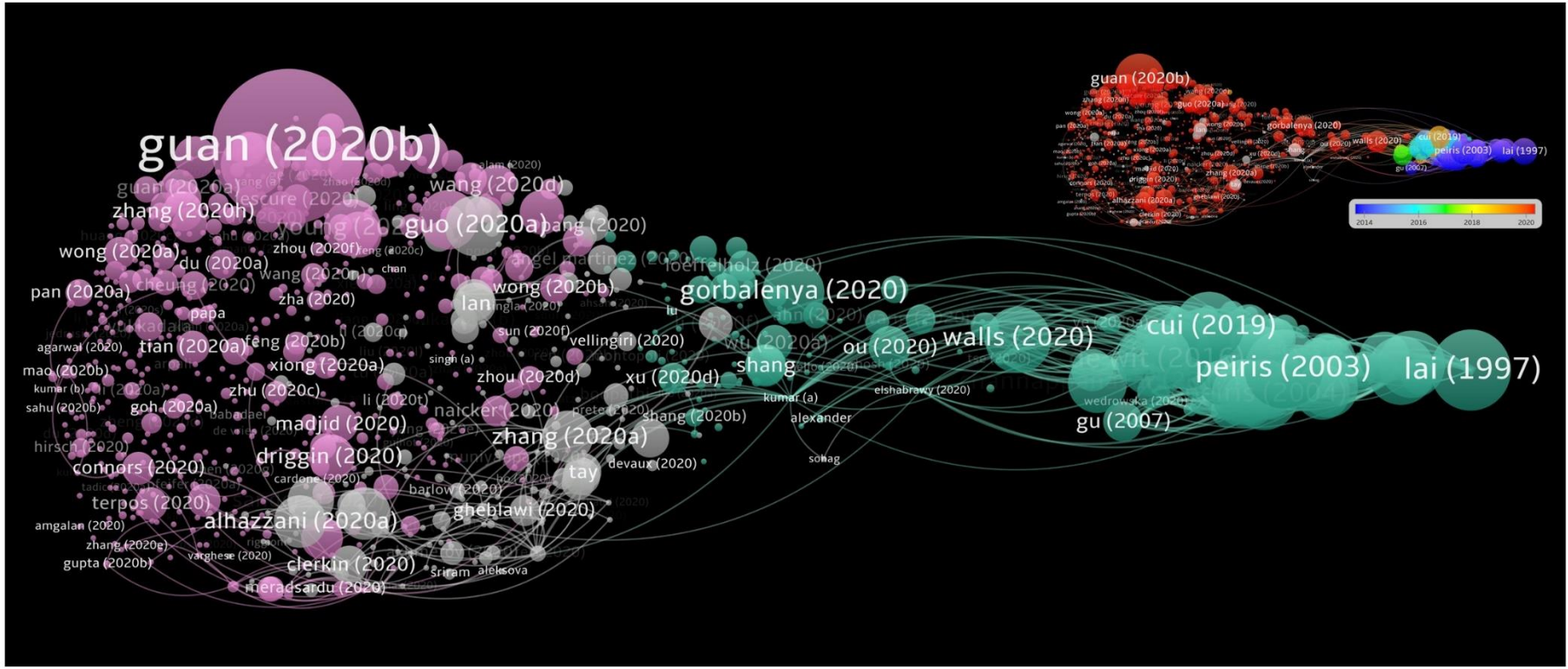

**Figure 2** Bibliographic coupling of coronavirus publications and an overlay with a counterpart map that visualises year of publication (top right corner).

*3.3. Co-citation network of the coronavirus literature*

The main metric on which we focus here in order to identify the streams of studies in the coronavirus literature is their co-citation relation. Two articles are co-cited when they jointly appear in the reference list of a third article. It is assumed that the higher the frequency of co-citation occurrences associated with two articles, the more similar they are in terms of content. In other words, two articles that are frequently cited simultaneously by other documents are deemed to have a strong co-citation relation, a standard and well-established sign of thematic similarity between articles (Small, 1973). A network of co-citation analysis visualises this relation between articles while identifying clusters of articles with strong co-citation relations. The network view of the co-citation map for thew coronavirus literature, consistent with the ground-breaking methodology of Chen (2004) for progressively visualising a knowledge domain, has been visualised in Figure 3. The node selection is based on the *g-index* algorithm (Chen, 2006). Each node represents a document while the size of the node is proportional to the number of local citations received by that document, i.e., the citations to that document exclusively from the studies of the coronavirus literature. A link between a pair of two nodes represents their co-citation relation, meaning that other documents of this literature have frequently co-cited the two articles in their references. Nodes with a red bubble are articles for which a citation burst has been detected. A burst of citation indicates a sudden increase in the number of citations to a document over a certain period of time. Here, bursts of citations to documents are detected based on the algorithm of Kleinberg (2003) adopted by CiteSpace (Chen, 2006; Chen et al., 2012). A burst is defined by its *start year*, *end year* (and as a result, its *duration*), as well as its *strength*. The size of a red bubble on the map for an article with a citation burst is an indication of the duration of its burst. In the Appendix, a list of sixty documents of the coronavirus literature with the longest citation bursts has been provided. A cluster with many citation bursts is indicative of a highly active area of research that may have been an emerging trend during a certain period of time.

In order to include a temporal component in the co-citation analysis, to visualise the progression of the knowledge development in this field, the co-citation algorithm divides the articles into time slices and constructs the temporal evolution of each cluster of studies in a timeline view. A video in the Online Supplementary Material (Supplementary Video #1) dynamically visualises the temporal evolution of this literature, based on its document co-citation network, over the last thirty years with time increments of one year. It can be observed from this supplementary video how vast parts of this literature have been constructed during only one year, that is, 2020 (or to be more precise, the first eight months of 2020). Figure 4 visualises segments of the network that have been constructed during these eight months of 2020; a reflection of co-citation instances (i.e., extent of activities) occurred during this period. Figure 5 visualises the equivalent timeline view of the map in Figure 3 with time slices of one year. In both maps, darker colours represent clusters that are relatively older while brighter colours (closer to yellow) represent younger clusters of references.

In order to facilitate presentation of the clusters and their content and nature, CiteSpace automatically assigns an objectively determined label to each cluster of co-cited documents, determined using a *loglikelihood ratio* (LR) algorithm applied to noun phrases extracted from the title of the citing articles (i.e. the articles that have co-cited articles of that cluster). This is meant to help interpret the contents of various clusters, although the labels need to be considered with an important caveat. It should be noted that a single label, per se, may not be an adequate representation of the clusters' content and nature. A cluster could embody various interconnected subtopics. Therefore, a list of top terms determined by the LR algorithm should be considered to gain more accurate insight into the content of each cluster. In other words, the label visualised on the map next to each cluster should merely be regarded as an abstract guide and not an absolute and literal description of the entire cluster content. In the visualised network, only major clusters receive a visualised label; therefore, there could be a discontinuity in the numbering of the clusters, especially as the clusters become smaller. A metric called *Silhouette Score* (SS) ([Lovmar et al., 2005](#)) is assigned to each cluster, whose value ranges theoretically between -1 and 1. This score is an indication of the quality of clustering with values closer to 1 indicative of more *homogenous* clusters. Another relevant metric is the *modularity Q* of the network ([Muff et al., 2005](#)), which measures the extent to which the network is decomposable into multiple components (here, clusters). The modularity Q of the network presented in Figure 3 is 0.8243, which is a relatively high value. The network is comprised of 4,383 nodes and 24,235 links.

In order to unpack details of the major clusters of co-cited articles in the coronavirus literature, Table 1 was constructed. In this table, for each major cluster, the *size* (i.e. number of documents associated with that cluster), the *mean year* of the publications associated with that cluster, the *silhouette score*, and a list of *top terms* extracted from the titles of the citing documents of that cluster (as determined based on the LR algorithm) are presented alongside the individual documents of each cluster that have strongest citation bursts and/or largest citation count. Documents within each cluster that show no burst but rank among top locally cited items have been included in the table as *influential articles*. The reason for this consideration is that these could be fundamental papers that have attracted a large number of local citations from within the field, while not having experienced any particular period of burst, rather, having accumulated their citations at a steady pace.

The timeline view of the co-citation network in Figure 5 clearly demonstrates the event-driven nature of the coronavirus literature with many of the highly locally cited articles being those published shortly after the SARS or MERS outbreaks. In fact, prior to 2020, the year following the outbreak of SARS, 2003, could be considered as a pivotal point in the development of this literature. Many of the articles of this literature with the highest local citations and strongest bursts are those published in the years immediately following the first SARS outbreak. In fact, activities represented by the majority of clusters of this literature seem to have initiated (or at least been invigorated) after this pivotal point, although there are clusters of research activities that existed prior to this time. Cluster #3 *neurotropic coronavirus*, for example, is one of the oldest and pioneering clusters of this field whose activities have now largely extinguished. Cluster #10, *subgenomic mRNA synthesis*, Cluster #20, *bovine coronavirus* and Cluster #23, *transmissible gastroenteritis virus* are other examples of clusters that existed before the first SARS outbreak, but seem to be no longer active. Cluster #12, *respiratory viruses*, is an example of a relatively short-lived stream of publications prompted by the SARS outbreak that had markedly reduced by the time that the MERS outbreak occurred (please see the Supplementary Video). However, the majority of clusters developed since 2003 seem to have continued their activities and are still ongoing to this date (e.g. Clusters #0, 2, 4, 7, 9, 16). The undelaying articles of these clusters (with the exception of cluster #7) seem to be heavily co-cited with 2020 articles, evidenced by the fact that there are links connecting various sections of these clusters (on the timeline view) to the items concentrated on the 2020 point of the timeline. This shows that the set of citing articles of these clusters include 2020 items too. Hence, these clusters still play a relevant role in the development of the SARS-CoV-2 literature. In other words, these streams of activity seem to be ongoing and are kept alive by the emerging studies related to the SARS-CoV-2 disease.

The map in Figure 5 also displays notable degrees of inter-cluster relations (i.e., notable co-citation occurrences between clusters, represented as vertical relations on the timeline) that coincide with the time of the occurrences of the three major coronavirus epidemics. This could also be another reflection of the heightened research activity at those points in time.

Certain clusters/streams of studies prompted by the MERS outbreak are also identifiable through the map of co-citation. This includes cluster #5 and cluster #13, both labelled *middle east respiratory syndrome coronavirus*. They both seem to represent streams of research that came to life in the wake of the MERS outbreak. Both clusters still show signs of activity to this date although less strongly compared to their onset.

Cluster #0, labelled *receptor-binding domain*, is also another cluster whose origin of activities coincides with the first major coronavirus outbreak (i.e., SARS). It represents one of the largest, most sustained and most fundamental streams of coronavirus research that is still relevant to the emerging trends of this field. Many fundamental articles of this cluster that have been published around 2003 (or shortly after)—the time of the first SARS epidemic—seem to have been heavily co-cited alongside 2020 articles of this field and by 2020 articles.

One of the oldest citation bursts of the coronavirus literature belongs to cluster #20 and is attributable to the study of [Sanger et al. (1977)](#) ("DNA sequencing with chain-terminating inhibitors", in *PNAS*), a highly fundamental article with nearly 68,000 total global citations (according to the WoS record) that has also attracted a considerable amount of attention from studies of the coronavirus literature, with a burst of citations in this particular field that lasted from 1990 to 1997. The strongest burst of citation in this literature has been recorded for the study of [Ksiazek et al. (2003)](#) ("A novel coronavirus associated with severe acute respiratory syndrome" in *New England Journal of Medicine*), an article that identifies with cluster #4, *severe acute respiratory syndrome*. This is followed by [Zaki et al. (2012)](#) (cluster #5), [Drosten et al. (2003)](#) (cluster #4), [Rota et al. (2003)](#) (cluster #4), [Peiris et al. (2003b)](#) (cluster #4) and [Marra et al. (2003)](#) (cluster #4).

Cluster #6 represents mostly research activities linked to the "*airborne transmissibility*" [Conly et al. (2020)](#) of the coronaviruses and the subtopics related to the use of "*face masks*", "*medical masks*", "*personal protective equipment*" and the protection of "*healthcare workers*". The visualisation of this cluster points to a number of articles prior to 2020 that are consistently referenced by the SARS-CoV-2 studies including the 2016 systematic review and meta analysis of [Radonovich et al. (2019)](#) as well as articles of [Guérin et al. (2013)](#), [Tran et al. (2012)](#) and [van der Sande et al. (2008)](#). Note that any link with one end in 2020 has to have established by 2020 citing articles. The existence of such links with one end in 2020 is most notable for clusters #0, #2, #6, #9, #16, #19, #21 and #22 meaning that SARS-CoV-2 studies are consistently referencing older items of these cluster.

Note that we have set a minimum of at least two years for recording bursts of citations. As a result, with the literature of SARS-CoV-2 being younger than one year, no burst of citation could be recorded for the items published following the SARS-CoV-2 outbreak.

**Figure 3** Network of document co-citation for the coronavirus literature.

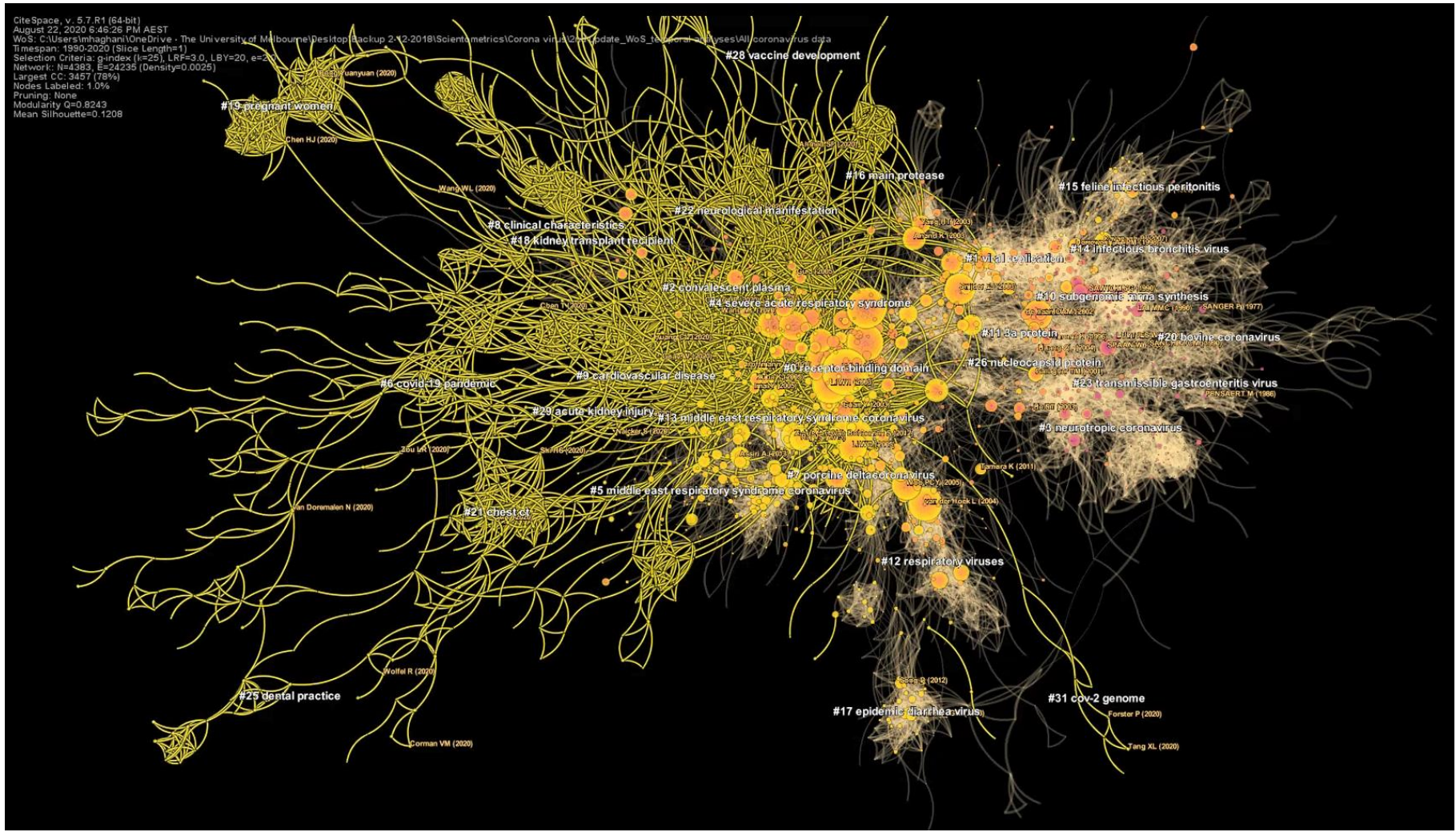

**Figure 4** The extent of research activities in coronavirus literature during first eight months of 2020, as reflected in the patterns of document co-citation. Visually Salient parts of the document co-citation map represent research activities (i.e., instances of document co-citation) during (first eight months of) 2020. See the dynamic visualisation in the Online Supplementary Video.

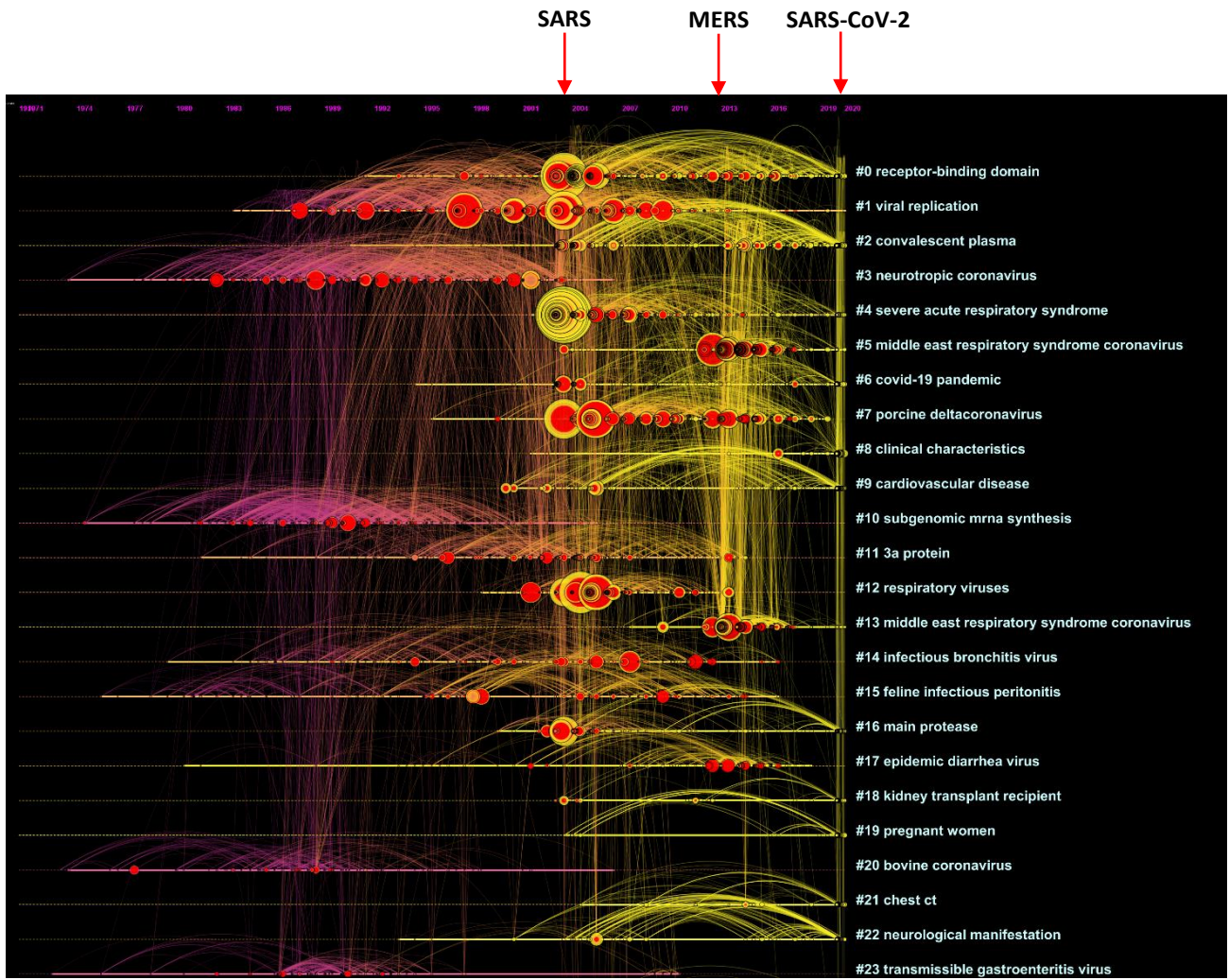

**Figure 5** Timeline view of the network of document co-citation for the coronavirus literature.

**Table 1** Cluster information for the document co-citation analysis of the coronavirus literature.

| ID | Size | Avg. year | SS | Top terms based on LL ratio | Cited articles with strongest citation bursts & Top cited articles |
|---|---|---|---|---|---|
| 0 | 255 | 2007 | 0.866 | **receptor-binding domain**; spike protein; potent neutralising antibodies; SARS coronavirus; human monoclonal antibodies; CoV-2 spike protein; human monoclonal SARS-CoV antibody; spike protein fusion core; binding domain; spike glycoprotein; viral entry; crystal structure; virus evolution | Li et al. (2003); Peiris et al. (2004); Bosch et al. (2008); Martina et al. (2003); Bisht et al. (2004); Sui et al. (2004); Xiao et al. (2003); Jeffers et al. (2004); Li et al. (2005a); Simmons et al. (2004); Li et al. (2005b); Belouzard et al. (2012); Li (2016); Hamming et al. (2004) |
| 1 | 246 | 2001 | 0.905 | **viral replication**; 3c-like proteinase; non-structural protein; coronavirus disease; coronavirus replication; virus replication; mouse hepatitis virus; Murine coronavirus; cleavage site; murine hepatitis virus; replicase gene polyprotein; RNA virus genome | Snijder et al. (2006); Lai and Cavanagh (1997); Thiel et al. (2003); Lee et al. (1991); Masters (2006); Perlman and Netland (2009); Ziebuhr et al. (2000); Gorbalenya et al. (2006); Knoops et al. (2008); Boursnell et al. (1987); Lai (2001); Cavanagh (1997) |
| 2 | 238 | 2013 | 0.882 | **convalescent plasma**; potential therapeutic option; potential therapy; clinical trial; synergistic effect; covid-19 treatment; clinical studies; emerging therapy | Al-Tawfiq et al. (2014); Cinatl et al. (2003a); So et al. (2003); Haagmans et al. (2004); Omrani et al. (2014); Cinatl et al. (2003b); Chan et al. (2013); Falzarano et al. (2013b); Poon et al. (2003); Dyall et al. (2014); De Wilde et al. (2014); Zumla et al. (2016); Chu et al. (2004) |
| 3 | 228 | 1992 | 0.924 | **neurotropic coronavirus**; central nervous system; mouse hepatitis virus; murine coronavirus spike protein; mouse hepatitis; murine coronavirus; spike glycoprotein; COVID-19 pandemic; virus-induced demyelination; amino acid substitution; soluble receptor | Spaan et al. (1988); Kuo et al. (2000); Wege and Ter Meulen (1982); Yeager et al. (1992); Dveksler et al. (1991); Delmas et al. (1992); Collins et al. (1982); Williams et al. (1991); Parker et al. (1989); Sturman et al. (1985); Fleming et al. (1986); Fuerst et al. (1986); Dalziel et al. (1986); Kubo et al. (1994); Gallagher and Buchmeier (2001) |
| 4 | 218 | 2008 | 0.830 | **severe acute respiratory syndrome**; papain-like protease; lymphocyte subset; antibody-dependent enhancement; coronavirus papain-like protease; non-structural protein; COVID-19 pathogenesis; clinical parameter; papain-like proteases; putative receptor; SARS-CoV2 infection; pneumonia patient | Ksiazek et al. (2003); Drosten et al. (2003); Rota et al. (2003); Peiris et al. (2003b); Marra et al. (2003); Lee et al. (2003); Peiris et al. (2003a); Poutanen et al. (2003); Tsang et al. (2003); Fouchier et al. (2003); Kuiken et al. (2003); Booth et al. (2003); Nicholls et al. (2003); Donnelly et al. (2003); Ruan et al. (2003); Bolles et al. (2011) |
| 5 | 187 | 2014 | 0.929 | **middle east respiratory syndrome coronavirus**; Saudi Arabia; dromedary camel; middle east respiratory syndrome; MERS coronavirus; middle east respiratory syndrome coronavirus infection; MERS-CoV infection; serological evidence; infected camel; Abu Dhabi | Zaki et al. (2012); Assiri et al. (2013b); Reusken et al. (2013); Haagmans et al. (2014); de Groot et al. (2013); Azhar et al. (2014); Memish et al. (2013a); Memish et al. (2013b); Guery et al. (2013); Oboho et al. (2015); Drosten et al. (2013) |
| 6 | 166 | 2011 | 0.971 | **COVID-19 pandemic**; face mask; personal protective equipment; healthcare worker; high-flow nasal cannula; airborne transmission; airway management; neck surgery; clinical practice; medical mask; surgical consideration | Riley et al. (2003); Lipsitch et al. (2003); Yu et al. (2004); Varia et al. (2003); Arabi et al. (2017); Dwosh et al. (2003); Twu et al. (2003); Seto et al. (2003); Alhazzani et al. (2020); Zou et al. (2020); Van Doremalen et al. (2020); Tran et al. (2012) |
| 7 | 164 | 2010 | 0.871 | **porcine delta-coronavirus**; genomic characterization; bat coronaviruses; interspecies transmission; human betacoronavirus; SARS-like coronaviruses; emerging viruses; metagenomic analysis; bat species | Guan et al. (2003); Li et al. (2005c); Lau et al. (2005); Woo et al. (2005); Ge et al. (2013); Annan et al. (2013); Ithete et al. (2013); Woo et al. (2009); Weiss and Navas-Martin (2005); Su et al. (2016); Fehr and Perlman (2015) |
| 8 | 152 | 2019 | 0.984 | **clinical characteristics**; novel coronavirus; middle east respiratory syndrome coronavirus; coronavirus disease; covid-19 patient; venous thromboembolism; cutaneous manifestation; retrospective study; clinical course; risk factor; clinical feature; epidemiological characteristics; Hubei province; COVID-19 infection | de Wit et al. (2016); Chan et al. (2020); Zhou et al. (2020) |

| 9 | 145 | 2016 | 0.962 | **cardiovascular disease**; angiotensin-converting enzyme; renin-angiotensin system inhibitor; angiotensin-converting enzyme inhibitor; cardiovascular system; angiotensin receptor blocker; heart failure; renin-angiotensin system | Tipnis et al. (2000); Kuba et al. (2005); Crackower et al. (2002); Imai et al. (2005); Donoghue et al. (2000); Harmer et al. (2002) |
|---|---|---|---|---|---|
| 10 | 145 | 1990 | 0.902 | **subgenomic mRNA synthesis**; mouse hepatitis virus; leader RNA; new transcriptional unit; expression system; ribonucleoprotein complex | Lai (1990); Sawicki and Sawicki (1990); Pachuk et al. (1989); Sethna et al. (1989) |
| 11 | 137 | 2001 | 0.913 | **3a protein; coronavirus disease**; severe acute respiratory syndrome coronavirus envelope protein; m protein; membrane protein; coronavirus pathogenesis | Vennema et al. (1996); Bosch et al. (2003); Almazán et al. (2013); Yount et al. (2005); Bos et al. (1996); Kuo and Masters (2003); Fischer et al. (1998); DeDiego et al. (2007) |
| 12 | 136 | 2005 | 0.970 | **respiratory viruses**; human coronavirus NL63; human bocavirus; acute respiratory tract infection; hematopoietic stem cell | van der Hoek et al. (2004); Woo et al. (2005); Fouchier et al. (2004); Peiris et al. (2004); Allander et al. (2005); Van den Hoogen et al. (2001); Lau et al. (2006) |
| 13 | 122 | 2014 | 0.937 | **middle east respiratory syndrome coronavirus**; mers-cov infection; MERS-CoV spike protein; COVID-19 pandemic; vaccine development; common marmoset; receptor-binding domain-based subunit vaccine; MERS-CoV vaccine | Raj et al. (2013); van Boheemen et al. (2012); Lu et al. (2013); Zhao et al. (2014); Wang et al. (2013); Falzarano et al. (2013a); De Wit et al. (2013); Müller et al. (2012); Du et al. (2009) |
| 14 | 107 | 1999 | 0.93 | **infectious bronchitis virus**; avian infectious bronchitis virus; turkey coronavirus; infectious bronchitis viruses; avian coronavirus; avian infectious bronchitis; genetic analysis; molecular analysis; RNA recombination | Tamura et al. (2011); Cavanagh (2007); Cavanagh (2005); Tamura et al. (2007); Thompson et al. (1994); De Wit et al. (2011); Cavanagh (2003); Cook et al. (2012) |
| 15 | 92 | 1997 | 0.974 | **feline infectious peritonitis**; feline coronavirus; feline infectious peritonitis virus; canine coronavirus | Pedersen et al. (2009); Herrewegh et al. (1998); Haijema et al. (2004) |
| 16 | 91 | 2011 | 0.972 | **main protease**; potential inhibitor; molecular docking; SARS-CoV-2 main protease; virtual screening; molecular dynamics | Anand et al. (2003); Yang et al. (2003); Anand et al. (2002); Fan et al. (2004); Wu et al. (2004) |
| 17 | 87 | 2010 | 0.979 | **epidemic diarrhea virus**; epidemic; porcine epidemic diarrhea virus | Stevenson et al. (2013); Song and Park (2012); Huang et al. (2013) |

*3.4. Co-citation network of the SARS-CoV-2 literature*

The size of the SARS-CoV-2 subset of the coronavirus literature is estimated to have exceeded 27,000 documents as of 22 August 2020 while exceedingly on the rise at the time of this analysis[1]. This subset, as recorded last time, is composed of nearly 43% Article, 22% Editorial, 20% Early Access, and 21% Letters among other document types. Four major categories in which documents of this literature have been indexed include *Medicine General Internal*, *Public Environmental Occupational Health*, *Infectious Diseases*, *Surgery and Immunology*. A relatively notable portion of this literature, nearly 8%, has been disseminated collectively across only six journals, out of the staggering number of nearly 4,780[2] journals that have published at least one document on this topic during the first eight months of 2020 (note that there are only a total of approximately 29,000 journal titles indexed by the WoS record). These major outlets of SARS-CoV-2 articles include *British Medical Journal* (766 items), *Journal of Medical Virology* (466 items), *Journal of Infection* (311 items), *Cureus* (244 items), *Lancet* (232 items) and *International Journal of Environmental Research and Public Health* (227 items). In 7,534 documents of this literature, at least one author affiliated with organisations of the United States of America has been listed. This is followed by authors affiliated with organisations of Peoples Republic of China (4,463 items) and Italy (3,332 items). The three organisations with the largest number of research items published on this subset of the coronavirus literature are *The Huazhong University of Science and Technology* (585 items), *Harvard Medical School* (475 items) and *Wuhan University* (356 items).

---

[1] A renewed search at the time of preparing the revision of this article, April 2021, shows that the size of the SARS-CoV-2 literature now stands at the staggering number of N=107,222 while the yearly split is N(2020)=83,208 and N(2021, till April)≈24,000.

[2] This number rose above 8,700 by April 2021, at the time of revising this article. The implication is that nearly a third of all scientific journals indexed by the WoS have now published at least one article related Covid-19 (SARS-CoV-2).

Here, further document co-citation analysis is conducted on the specific subset of the SARS-CoV-2 literature with the aim of identifying documents published prior to 2020 (i.e., prior to the outbreak of SARS-CoV-2) that have proven instrumental in informing the 2020 SARS-CoV-2-related articles. In other words, the main objective of the analysis is to determine documents outside the subset of SARS-CoV-2 literature (i.e., those published prior to 2020) that are frequently co-cited with SARS-CoV-2-related documents and by SARS-CoV-2-related documents. A timeline view of the document co-citation network associated with this literature, exclusive to the major clusters only, is presented in Figure 6. This network has 966 nodes and 6,347 links and has a modularity Q of 0.574. As expected, the majority of the network congregates on the 2020 cross section of the timeline. However, almost every major cluster displays a number of older documents frequently co-cited with the 2020 articles of that cluster, as evidenced by the links going back from 2020 to the previous years on the timeline. Note that, here, every citing article is a 2020 SARS-CoV-2 article (with exception of a very few 2019 and very few 2021 items). Therefore, every link between 2020 and a point prior to 2020 is indicative of the existence of an article published prior to 2020 (i.e., outside the SARS-CoV-2 dataset) that has been jointly cited with a specific article of SARS-CoV-2 literature, and by several other articles of the SARS-CoV-2 literature.

Some clusters such as cluster #1, cluster #3 and cluster #4 look back at instrumental documents dating around 2004 and 2005; and some other clusters look back at more recent articles. This includes cluster #7 and cluster#8, for example. Cluster #9 does not look back to any prominent document within that cluster, but instead, it refers back a particular 2004 document (Hamming et al., 2004) affiliated with another cluster, cluster #3. Note that, in interpreting this timeline view map, a link between Article A (2020) and Article B (<2020) does not necessarily indicate that Article A has looked back at B and has cited it, rather, it means that there have been "a number of" other Articles like C (all 2020) that have looked back at Article B in their development and have it (alongside Article A).

Among the documents published prior to 2020 that are frequently co-cited with SARS-CoV-2 studies, two items seem to have been particularly instrumental considering the number of local citations that they have received from the SARS-CoV-2 papers. One is the review article of Tran et al. (2012) ("Aerosol generating procedures and risk of transmission of acute respiratory infections to healthcare workers: a systematic review" in *PLOS ONE*) that has been frequently co-cited with 2020 studies of cluster #2 labelled "*oral health*", also to lesser degrees, with those of cluster #0, "*novel coronavirus*". The other highly influential item is a 2017 document entitled "*Community Mitigation Guidelines to Prevent Pandemic Influenza — United States, 2017*" (Qualls et al., 2017) that has been frequently cited in conjunction with 2020 studies of cluster #8, "*mental health*", also to lesser degrees, with those of cluster #0.

In order to unpack further details of the SARS-CoV-2 document co-citation network, information related to the major clusters of this network have been synthesised in Table 2. The last column of the table singles out the items of the SARS-Cov-2 literature (i.e., the 2020 items) that have, thus far, received largest local citation count from within that literature, as well as most prominent documents of each cluster published prior to 2020, i.e. items outside the domain of SARS-CoV-2 literature. These items have been marked by a † sign in the table.

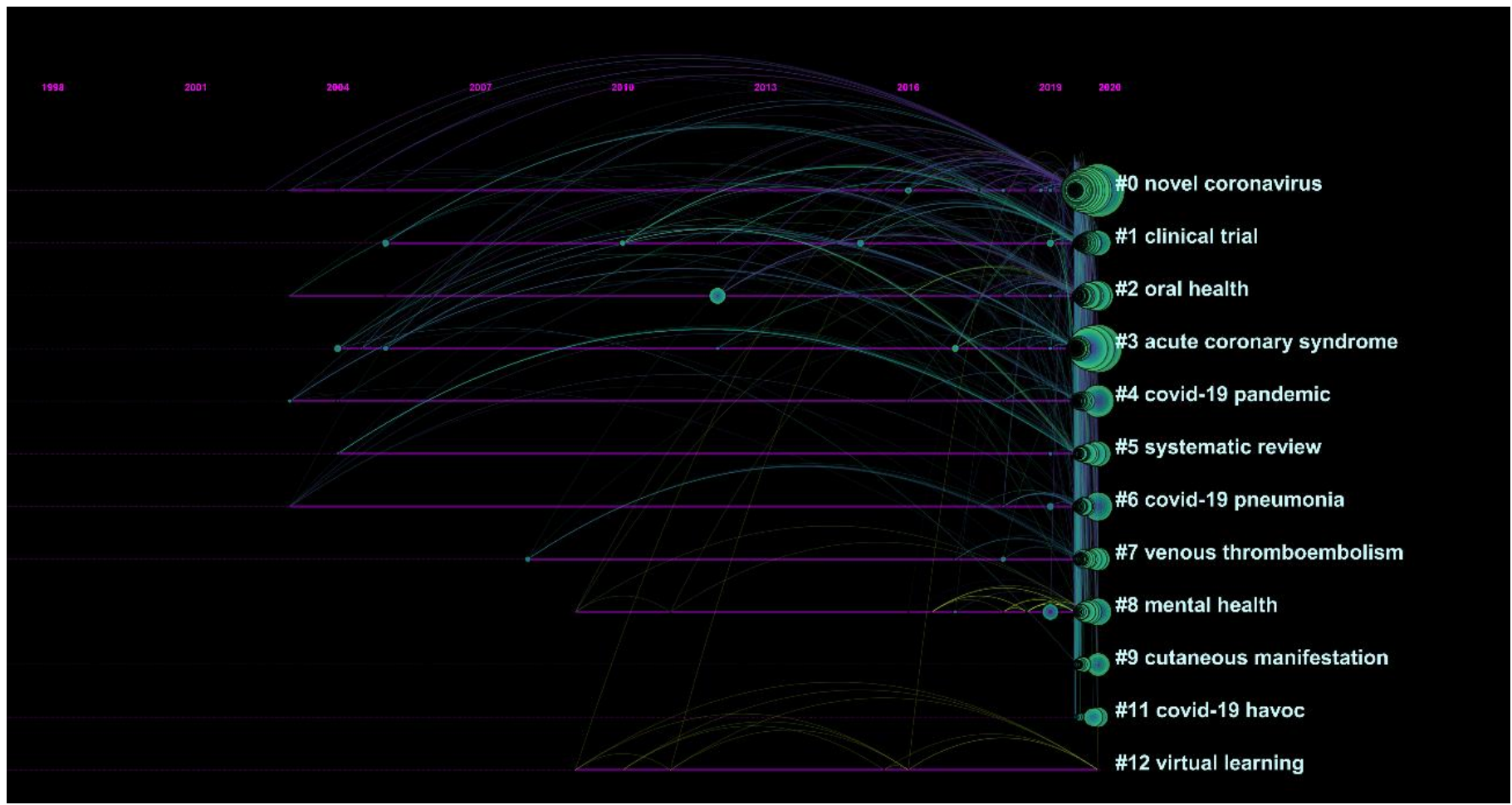

**Figure 6** Timeline view of the document co-citation network of the SARS-CoV-2 literature.

**Table 2** Cluster information of the document co-citation analysis of the SARS-CoV-2 literature.

| ID | Size | SS | Top terms based on LL ratio | Cited articles with largest local citation counts + items published prior to 2020 with notable local citation count (marked by †) |
|---|---|---|---|---|
| 0 | 129 | 0.751 | **novel coronavirus**; new coronavirus; novel coronavirus outbreak; laboratory finding | Huang et al. (2020); Wang et al. (2020a); Zhu et al. (2020b); Su et al. (2016)† |
| 1 | 117 | 0.751 | **clinical trial**; lung transplant recipient; solid organ transplant recipient; convalescent plasma; kidney transplantation; severe COVID-19 pneumonia; renal transplant recipient; effective cure; anti COVID-19 drug; pre-clinical data | Mehta et al. (2020); Wang et al. (2020b); Gautret et al. (2020); Cao et al. (2020); Dong et al. (2020a); Wu et al. (2020b); Rothan and Byrareddy (2020); Alhazzani et al. (2020); Lai et al. (2020); Moher (2010)†; Fehr and Perlman (2015)†; Vincent et al. (2005)†; Carbajo-Lozoya et al. (2012)†; Mair-Jenkins et al. (2015)† |
| 2 | 113 | 0.776 | **oral health**; dental school; dental care; medical mask; triaging patient; endoscopy workflow; infection control practice; CoV-2 testing; virological diagnosis | Van Doremalen et al. (2020); Zou et al. (2020); Remuzzi and Remuzzi (2020); Wang et al. (2020d); Bai et al. (2020); Wölfel et al. (2020); Li et al. (2020b); Tran et al. (2012)†; Tellier et al. (2019)†; Judson and Munster (2019)†; Li et al. (2003)†; Harris et al. (2009)†; Stockman et al. (2006)† |
| 3 | 107 | 0.741 | **acute coronary syndrome**; COVID-19 severity; liver injury; myocardial infarction; cardiovascular magnetic resonance; cardiovascular disease; lung cancer; case-fatality rate | Guan et al. (2020); Wu et al. (2020c); Wu and McGoogan (2020); Yang et al. (2020b); Grasselli et al. (2020); Onder et al. (2020); Channappanavar and Perlman (2017)†; Hamming et al. (2004)†; Kuba et al. (2005)†; Imai et al. (2005)†; Kwong et al. (2018)† |
| 4 | 94 | 0.760 | **COVID-19 pandemic**; COVID-19 outbreak; cancer patient; surgical treatment; transplant surgery; neurosurgical practice; telemedicine | Liang et al. (2020); Spinelli and Pellino (2020); Collaborative (2020); Emanuel et al. (2020); Hollander and Carr (2020); Wei et al. (2003)†; Bray et al. (2018)†; Kwak et al. (2016)† |
| 5 | 78 | 0.858 | **systematic review**; retrospective review; paediatric patient; CoV-2 infection; COVID-19 in children; European resuscitation council; COVID-19 guideline; breast feeding; pregnancy; vertical transmission | Chen et al. (2020a); Dong et al. (2020c); Lu et al. (2020); Zhu et al. (2020a); Ludvigsson (2020b); Dong et al. (2020b); Radonovich et al. (2019)†; Wong et al. (2004)† |
| 6 | 76 | 0.877 | **COVID-19 pneumonia**; chest CT; lung ultrasound; patient management; Fleischer society; radiology department; monitoring disease | Ai et al. (2020); Pan et al. (2020); Bernheim et al. (2020); Chung et al. (2020); Fang et al. (2020); Shi et al. (2020); Koo et al. (2018)†; Gralinski et al. (2018)†; Ksiazek et al. (2003)† |
| 7 | 75 | 0.886 | **venous thromboembolism**; covid-19 infection; taste disorder; COVID-19 coagulopathy; pulmonary embolism; neurological complication; neurological manifestation | Tang et al. (2020b); Mao et al. (2020a); Tang et al. (2020a); Klok et al. (2020); Zhang et al. (2020d); Lechien et al. (2020); Lurie and Carr (2018)†; Netland et al. (2008)†; Kim et al. (2017)† |
| 8 | 71 | 0.779 | **mental health**; general population; eco-epidemiological assessment; dialysis facilities; hotline service | Brooks et al. (2020); Qualls et al. (2017)†; Al-Ahmadi et al. (2019)† |
| 9 | 34 | 0.947 | **cutaneous manifestation**; dermatological finding; cutaneous sign | Recalcati (2020); Galván Casas et al. (2020); Joob and Wiwanitkit (2020) |

† Marks articles published prior to SARS-COV-2 epidemic that have been frequently cited by publications on SARS-COV-2.

## 3.5. Sleeping beauties of the coronavirus literature

The term *sleeping beauty* in reference to scientific publications is a metaphoric term describing publications with delayed recognition, those that go relatively unnoticed or unrecognised for several years, but suddenly gain a high level of attention, later in their lifetime (van Raan, 2004). This determination is generally made based on the common metric of the number of citations recorded to a publication. Sleeping beauties are often, though not invariably, created as a result of an important scientific discovery or a major non-scientific event, a trigger or a cause that is usually referred to as the *prince* corresponding to that sleeping beauty, whose determination is not necessarily easy to make.

The previous co-citation analysis of the documents related to the SARS-CoV-2 disease, as a literature that has entirely developed within less than a year at the time of this analysis, revealed a certain set of articles outside the specific SARS-CoV-2 dataset, all published prior to 2020, that appear to have become consistently contributory to the intellectual development of the studies emerged in the wake of the 2020 disease outbreak, in that, they are consistently and frequently cited by SARS-CoV-2 papers. This is an indication that this set of articles are playing an important role in shaping the rapidly-evolving SARS-CoV-2 literature, and hence, could be candidates for being sleeping beauties of the coronavirus literature, in which case and following the same line of metaphoric language, the SARS-CoV-2 pandemic should be deemed as the prince that is causing their awakening.

To investigate this question, articles marked in Table 2 by a † sign, were individually investigated and their history of citations (as recorded by WoS) were one by one analysed. The investigation showed that a large portion of such studies, though not all of them, were in fact sleeping beauties of this field. These publications have been listed in Table 3 with their respective citation history is visualised in Figure 8. Clearly, not all marked articles of Table 2 qualified as a sleeping beauty. For example, we required that the study has at least three years of citation history. Therefore, articles published in 2019, 2018 or 2017 were automatically left out of the consideration. Also, studies whose record shows an increasing trend of citation accumulation were not included in this list, even if the number of citations to them in 2020 was substantially higher compared to the earlier years. An example of this is the article of Moher (2010), that has been well and truly awake and accumulating increasing number of citations since birth, even though it has been cited at a much higher rate in 2020. Figure 7 illustrates this concept and further clarifies our method of identifying sleeping beauty candidates.

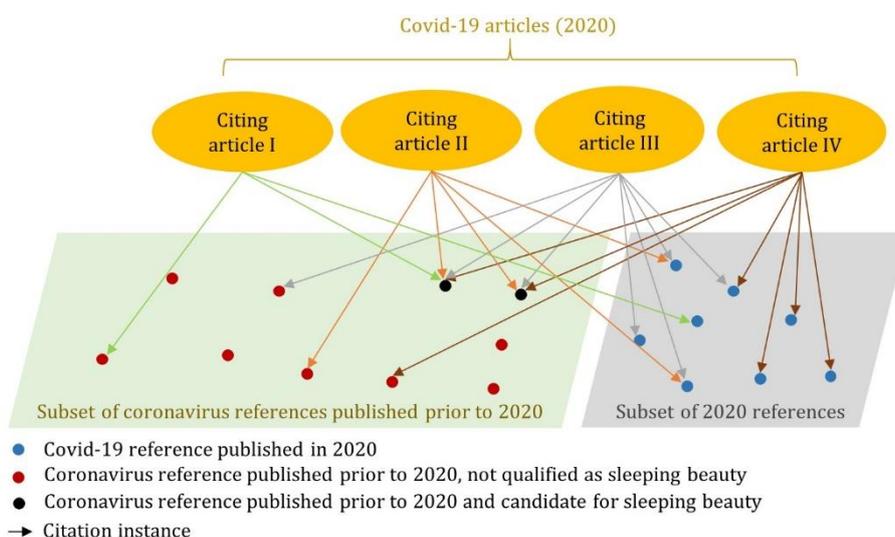

**Figure 7** Illustration of the method used for identifying potential sleeping beauties of the coronavirus literature, exclusive to those that had an awakening in 2020.

Some of the articles identified in this list may not have been absolute "sleeping" items in the lifetime, rather, there were a few that had a relatively notable amount of attention in their earlier years while having lost traction shortly after the first round of attention and remaining relatively unnoticed over the years leading to 2020, before becoming a centre of attention again in 2020. This includes articles such as those of Mair-Jenkins et al. (2015), and Li et al. (2003).

Our search process also identified a number of articles published within the range of 2017-2019, that cannot technically qualify as sleeping beauties of this literature (as they are too recent) but have gained exceptional attention from the 2020 SARS-CoV-2 studies. One can mention the study of Lurie and Carr (2018) as a prime example. The article is titled "The role of telehealth in the medical response to disasters" and is a short Viewpoint of only two pages long published by JAMA Internal Medicine in April 2018. The article had not received a single citation during 2018 or 2019, and yet, was poised to become an influential document in the wake of the novel coronavirus epidemic. Its authors, Lurie and Carr, had been so visionary to recognise the importance of "technologies to support and promote long-distance clinical care, education and health administration" in the events of disasters that could "cause damage, ecological disruption, loss of human life, or deterioration of health and health services", an idea that could not have been any timelier and more relevant once the SARS-CoV-2 pandemic affected the world. A look at the studies that have so far cited this article demonstrates how the ideas discussed by these authors two years ahead of a major global health emergency are now proving to be critically important. Other examples of similar studies that, due to their young age, could not be classified as sleeping beauties, but are receiving exceptional attention (relative to their own (short) history) from the SARS-CoV-2 papers are Channappanavar and Perlman (2017), Kwak et al. (2016), Gralinski et al. (2018) and Kim et al. (2017).

Note that our analysis in this section should not be considered as an attempt to exhaustively identify all possible sleeping beauties across the entire coronavirus literature. Rather, the analysis focused particularly on identifying the recently awakened items, as opposed to possible sleeping beauties that may have had an awakening earlier by other prince(s) such as SARS or MERS epidemics. That remains beyond the scope of our analysis.

**Table 3** Sleeping beauties of coronavirus literature with at least five years sleep, awakened in 2020.

| Title | Authors (year) | Journal |
|---|---|---|
| Pregnancy and perinatal outcomes of women with severe acute respiratory syndrome | Wong et al. (2004) | American Journal of Obstetrics and Gynecology |
| Coronaviruses: An overview of their replication and pathogenesis | Fehr and Perlman (2015) | N.A. |
| Chloroquine is a potent inhibitor of SARS coronavirus infection and spread | Vincent et al. (2005) | Virology Journal |
| Replication of human coronaviruses SARS-CoV, HCoV-NL63 and HCoV-229E is inhibited by the drug FK506 | Carbajo-Lozoya et al. (2012) | Virus Research |
| The Effectiveness of Convalescent Plasma and Hyperimmune Immunoglobulin for the Treatment of Severe Acute Respiratory Infections of Viral Etiology: A Systematic Review and Exploratory Meta-analysis | Mair-Jenkins et al. (2015) | Journal of Infectious Diseases |
| Aerosol generating procedures and risk of transmission of acute respiratory infections to healthcare workers: a systematic review | Tran et al. (2012) | PLOS ONE |
| Angiotensin-converting enzyme 2 is a functional receptor for the SARS coronavirus | Li et al. (2003) | Nature |
| SARS: systematic review of treatment effects | Stockman et al. (2006) | PLoS Medicine |
| A Double-Inactivated Severe Acute Respiratory Syndrome Coronavirus Vaccine Provides Incomplete Protection in Mice and Induces Increased Eosinophilic Proinflammatory Pulmonary Response. | Bolles et al. (2011) | Journal of Virology |
| Tissue distribution of ACE2 protein, the functional receptor for SARS coronavirus. A first step in understanding SARS pathogenesis | Hamming et al. (2004) | Journal of Pathology |
| A crucial role of angiotensin converting enzyme 2 (ACE2) in SARS coronavirus-induced lung injury | Kuba et al. (2005) | Nature Medicine |
| Angiotensin-converting enzyme 2 protects from severe acute lung failure | Imai et al. (2005) | Nature |
| Safe tracheostomy for patients with severe acute respiratory syndrome | Wei et al. (2003) | The Laryngoscope |
| Severe acute respiratory syndrome coronavirus infection causes neuronal death in the absence of encephalitis in mice transgenic for human ACE2 | Netland et al. (2008) | Journal of Virology |
| Professional and Home-Made Face Masks Reduce Exposure to Respiratory Infections among the General Population | van der Sande et al. (2008) | PLOS ONE |

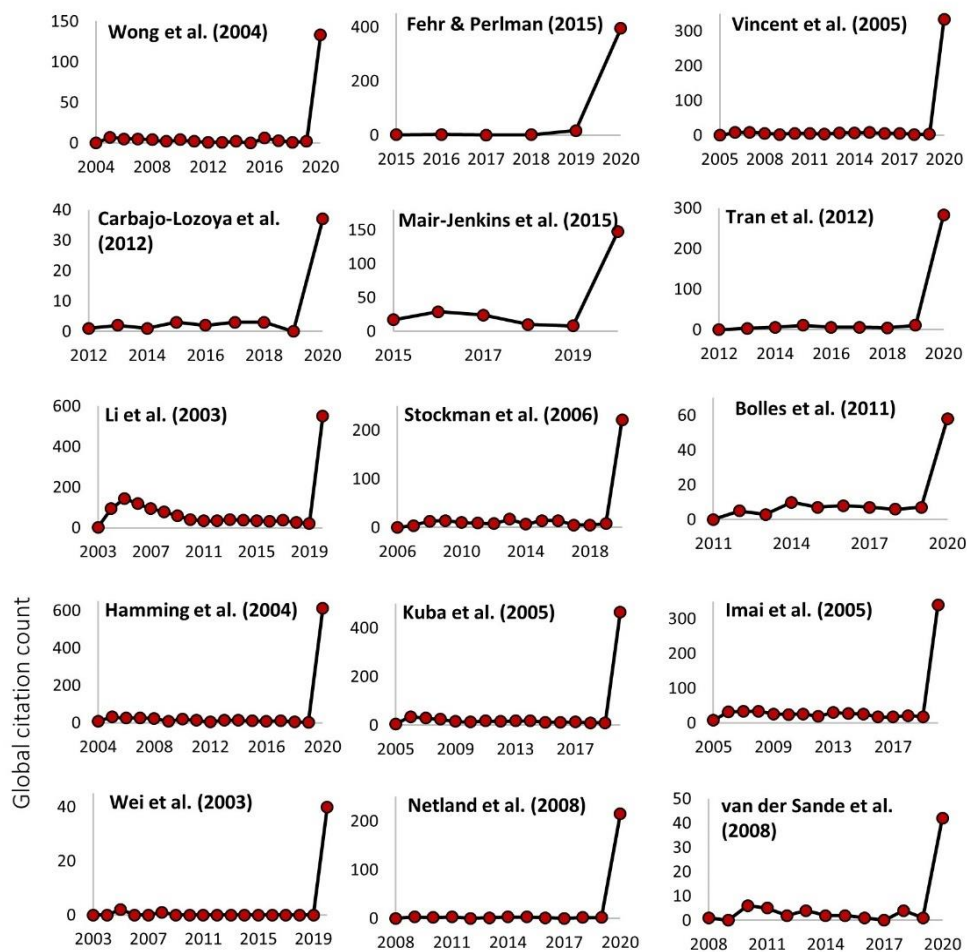

**Figure 8** Citation history of the sleeping beauties of the coronavirus literature, those with an awakening in 2020 and at least five-years-long sleep period.

*3.6. Reviews and meta analyses of the coronavirus literature*

As of September 2020, more than 1,000 review and meta studies have been recorded by WoS related to the coronavirus literature (1,024 items on 4 September 2020)[3], of which nearly 88% (902 items) are related to SARS-CoV-2 and have been published in 2020[4]. These reviews and meta analyses have been distributed across more than 650 journals/sources with *Journal of Medical Virology* (48 items), *Cureus* (37 items) and *Medicine* (29 items) being major outlets for these reviews and meta studies. Table 4 lists the top 30 most cited reviews of the coronavirus literature based on their total citation count recorded by WoS. Of these items, more than half are documents published since January 2020.

**Table 4** Top 30 most cited review and meta studies of the coronavirus literature.

| | Title | Authors (year) | Journal | Citation count |
|---|---|---|---|---|
| 1. | The psychological impact of quarantine and how to reduce it: rapid review of the evidence | Brooks et al. (2020) | Lancet | 500 |
| 2. | Porcine epidemic diarrhoea virus: a comprehensive review of molecular epidemiology, diagnosis, and vaccines | Song and Park (2012) | Virus genes | 355 |
| 3. | Role of ventilation in airborne transmission of infectious agents in the built environment - a multidisciplinary systematic review | Li et al. (2007) | Indoor Air | 349 |
| 4. | SARS: Systematic review of treatment effects | Stockman et al. (2006) | PLOS Medicine | 346 |
| 5. | Incubation periods of acute respiratory viral infections: a systematic review | Lessler et al. (2009) | Lancet Infectious Diseases | 299 |
| 6. | Prevalence of comorbidities and its effects in patients infected with SARS-CoV-2: a systematic review and meta-analysis | Yang et al. (2020a) | International Journal of Infectious Diseases | 263 |
| 7. | World Health Organization declares global emergency: A review of the 2019 novel coronavirus (COVID-19) | Sohrabi et al. (2020) | International Journal of Surgery | 263 |
| 8. | The Effectiveness of Convalescent Plasma and Hyperimmune Immunoglobulin for the Treatment of Severe Acute Respiratory Infections of Viral Etiology: A Systematic Review and Exploratory Meta-analysis | Mair-Jenkins et al. (2015) | Journal of Infectious Diseases | 222 |
| 9. | A review of feline infectious peritonitis virus infection: 1963-2008 | Pedersen (2009) | Journal of Feline Medicine and Surgery | 202 |
| 10. | Pharmacologic Treatments for Coronavirus Disease 2019 (COVID-19) A Review | Sanders et al. (2020) | Jama-Journal of the American Medical Association | 187 |
| 11. | Clinical, laboratory and imaging features of COVID-19: A systematic review and meta-analysis | Rodriguez-Morales et al. (2020) | Travel Medicine and Infectious Disease | 173 |
| 12. | A review of viral gastroenteritis | Clark and McKendrick (2004) | Current Opinion in Infectious Diseases | 170 |
| 13. | A Review of Coronavirus Disease-2019 (COVID-19) | Singhal (2020) | Indian Journal of Pediatrics | 153 |
| 14. | Systematic review of COVID-19 in children shows milder cases and a better prognosis than adults | Ludvigsson (2020a) | Acta Paediatrica | 150 |
| 15. | Epidemiology, causes, clinical manifestation and diagnosis, prevention and control of coronavirus disease (COVID-19) during the early outbreak period: a scoping review | Adhikari et al. (2020) | Infectious Diseases of Poverty | 140 |
| 16. | Viruses and bacteria in acute asthma exacerbations - A GA2LEN-DARE* systematic review | Papadopoulos et al. (2011) | Allergy | 137 |
| 17. | A systematic review on the efficacy and safety of chloroquine for the treatment of COVID-19 | Cortegiani et al. (2020) | Journal of Critical Care | 136 |
| 18. | Thrombocytopenia is associated with severe coronavirus disease 2019 (COVID-19) infections: A meta-analysis | Lippi et al. (2020) | Clinica Chimica Acta | 128 |
| 19. | Demographic and attitudinal determinants of protective behaviours during a pandemic: A review | Bish and Michie (2010) | British Journal of Health Psychology | 125 |
| 20. | Potential interventions for novel coronavirus in China: A systematic review | Zhang and Liu (2020) | Journal of Medical Virology | 124 |
| 21. | Coronavirus Disease 2019 (COVID-19): A Systematic Review of Imaging Findings in 919 Patients | Salehi et al. (2020) | American Journal of Roentgenology | 116 |
| 22. | Potential Effects of Coronaviruses on the Cardiovascular System A Review | Madjid et al. (2020) | Jama Cardiology | 112 |
| 23. | Prevalence of comorbidities in the Middle East respiratory syndrome coronavirus (MERS-CoV): a systematic review and meta-analysis | Badawi and Ryoo (2016) | International Journal of Infectious Diseases | 105 |
| 24. | Porcine epidemic diarrhea: a review of current epidemiology and available vaccines | Song et al. (2015) | Clinical and Experimental Vaccine Research | 105 |

---

[3] This number, in April 2021, when the revision of this article was being prepared had risen to N=3586, of which N(2020)=2,642 and N(2021, till April)≈820, according to the WoS.

[4] The implication of the updated figures (obtained in April 2021) is that the share of the review articles of coronavirus literature published post Covid-19 pandemic, relative to the entire size of the review articles in this literature, neared 97%, in April 2021 (an increase from 88% in September 2020, and a clear sign of increasing accumulation of review articles in this field).

| 25. | A review of vaccine research and development: Human acute respiratory infections | Girard et al. (2005) | Vaccine | 105 |
| 26. | A review of studies on animal reservoirs of the SARS coronavirus | Shi and Hu (2008) | Virus Research | 103 |
| 27. | COVID-19 and smoking: A systematic review of the evidence | Vardavas and Nikitara (2020) | Tobacco Induced Diseases | 98 |
| 28. | COVID-19 patients' clinical characteristics, discharge rate, and fatality rate of meta-analysis | Li et al. (2020a) | Journal of Medical Virology | 92 |
| 29. | A review of feline infectious peritonitis virus - molecular-biology, immunopathogenesis, clinical aspects, and vaccination | Olsen (1993) | Veterinary Microbiology | 91 |
| 30. | Hypothesis for potential pathogenesis of SARS-CoV-2 infection-a review of immune changes in patients with viral pneumonia | Lin et al. (2020) | Emerging Microbes & Infections | 89 |

Figure 9 presents a hybrid map of bibliographic coupling of the review and meta studies of the coronavirus literature. This figure also particularly allows us to contrast the extent of research synthesis activities in relation to the coronavirus literature during the first eight months of 2020 as opposed to the years before. The colour yellow has been assigned to mark the 2020 items and to distinguish them from the rest of the body of review and meta studies in this literature. The map shows a coupling pattern comparable to that of the general literature of coronaviruses (Figure 2) by presenting the body of 2020 studies largely separate from the previous literature in terms of their number of shared references. Recent research synthesis studies (i.e., 2020 items) that show coupling relations with both subsets of the literature (i.e., between reviews and meta analyses published prior and after 2020) include Zhang and Liu (2020) and Pormohammad et al. (2020). The review article of Mo and Fisher (2016) is also one that has been prompted by the MERS outbreak and has been published at the early stages of that outbreak but is bibliographically linked to several of the individual 2020 review and meta analyses on SARS-CoV-2. This means that several of the 2020 review studies have a considerable number of references in common with those of Mo and Fisher (2016). Within the body of the 2020 review and meta-analytic articles, Kulkarni et al. (2020), Mao et al. (2020b), Sanders et al. (2020), Cheung et al. (2020), Kim et al. (2020), Zhang et al. (2020b) and Liguoro et al. (2020) are those with the strongest degree of bibliographic connectivity with the rest of the 2020 cluster, meaning that they each have a considerable number of references in common with several number of other 2020 review articles.

In order to identify the studies that have most frequently been cited by reviews and meta analyses of the coronavirus literature, and to determine the reviews and meta analyses with the highest degree of *coverage* within each cluster (i.e., to determine the citing reviews and meta analyses that have cited the largest number of items/references within each cluster), a document co-citation analysis was conducted on the sub-set of 1,024 review and meta-analytic articles. The map of document co-citation associated with this analysis overlaid with convex hulls that highlight the borders of each cluster has been shown in Figure 10. The network includes 1,229 nodes and 3,941 links and has a modularity Q of 0.921. Information associated with individual major clusters has been synthesised in Table 5. This includes, for each cluster, articles that have received largest local citation counts from reviews and meta analyses of this literature, as well as citing articles (i.e., reviews and meta analyses) with the largest degree of coverage. For some clusters such as cluster #0, cluster #1 and cluster #4, both major citing articles and major cited items are 2020 publications, an indication that the cluster is exclusively focused on SARS-CoV-2. The dynamic evolution of this map on a year-by-year basis can be more tangibly observed through the video associated with this map that is accessible in the Online Supplementary Material (Supplementary Video #2). The visually salient parts of this map (with thicker links) highlight particular segments/areas of the network that has been constructed since January 2020, in contrast to the rest of the network that has formed prior to the SARS-CoV-2 outbreak.

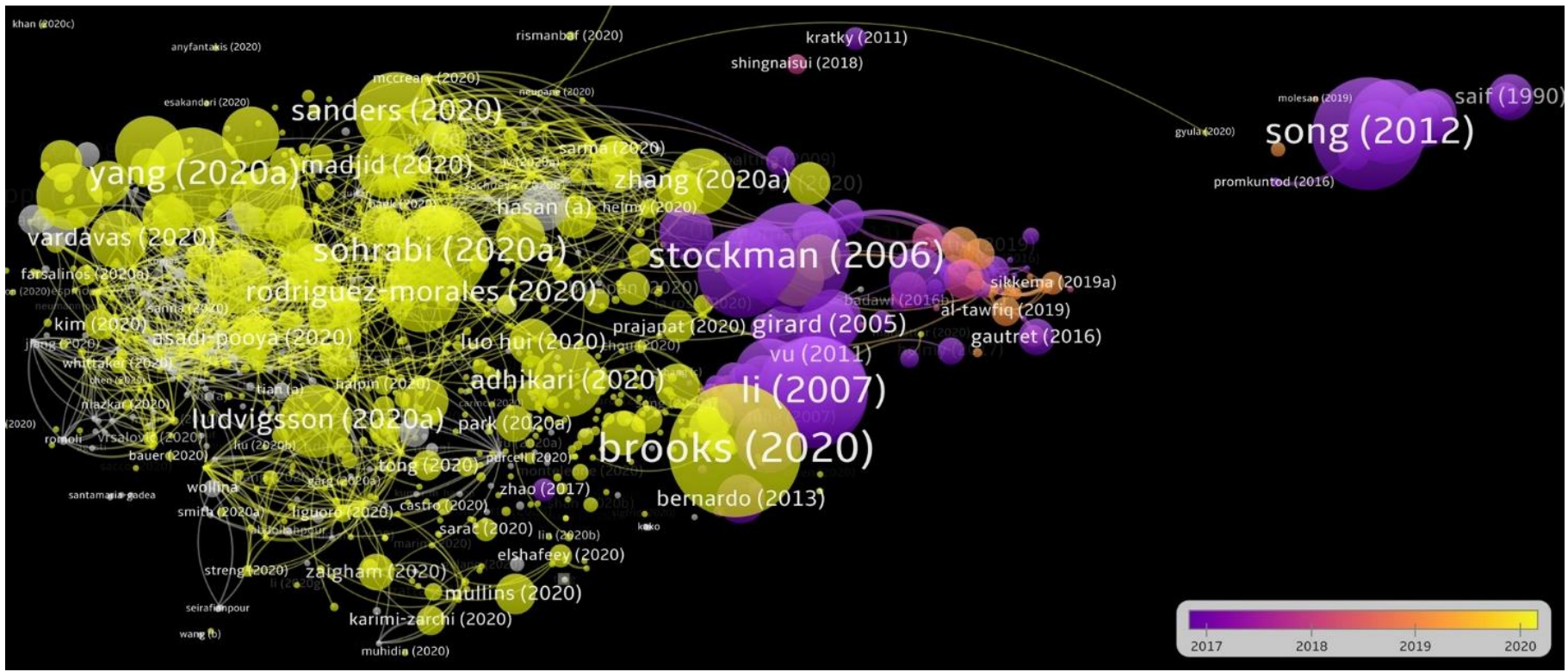

**Figure 9** Hybrid map of the bibliographic coupling of reviews and meta analyses in the coronavirus literature. The colour coding represents the year of publication.

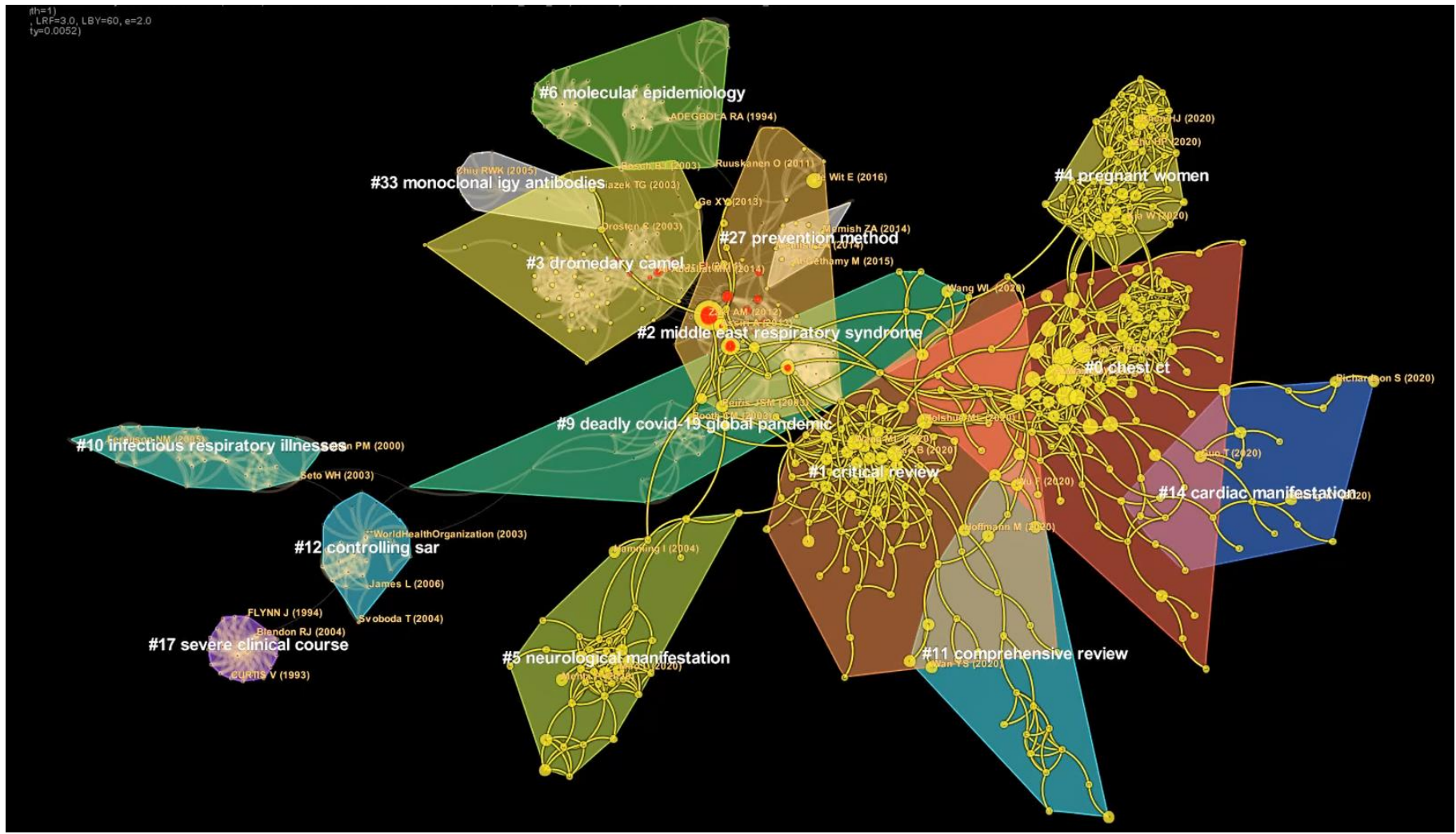

**Figure 10** Map of document co-citation for the subset of reviews and meta analyses in the general coronavirus Literature. Visually salient links represent co-citation instances occurred in 2020 (i.e., by 2020 reviews and meta analytical studies related to SARS-CoV-2).

**Table 5** Cluster information of the co-citation analysis of the coronavirus review and meta-analytic papers.

| Cluster id | Size | Silhouette score | Mean year | Top terms based on LL ratio | Items with largest local citation count | Citing articles with highest coverage |
|---|---|---|---|---|---|---|
| 0 | 100 | 0.951 | 2020 | **chest CT**; COVID-19 pneumonia; adult patient; chest CT finding; tomography finding; COVID-19 clinical | Huang et al. (2020); Wang et al. (2020a); Fu et al. (2020); Chen et al. (2020b) | Kulkarni et al. (2020); Yu and Chen (2020); Ortiz-Prado et al. (2020); Sun et al. (2020) |
| 1 | 81 | 0.981 | 2016 | **critical review**; treatment consideration; emerging option; pharmacologic treatment | Fu et al. (2020); Wang et al. (2020b); Cao et al. (2020) | Yu and Chen (2020); Ortiz-Prado et al. (2020); Neerukonda and Katneni (2020) |
| 2 | 63 | 0.948 | 2013 | **middle east respiratory syndrome**; middle east respiratory syndrome coronavirus; treatment modalities | Zaki et al. (2012); de Wit et al. (2016); Assiri et al. (2013a) | Zhu et al. (2019); Dawson et al. (2019); Gulati et al. (2020) |
| 3 | 63 | 0.955 | 2013 | **dromedary camel**; middle east respiratory syndrome coronavirus; MERS-CoV seroprevalence; RNA prevalence; zoonotic pathogen | Ge et al. (2013); Al-Abdallat et al. (2014); Azhar et al. (2014) | Sikkema et al. (2019); Dighe et al. (2019); Zhu et al. (2019) |
| 4 | 61 | 0.998 | 2019 | **pregnant women**; vertical transmission; clinical characteristics | Chen et al. (2020a); Zhu et al. (2020a); Xia et al. (2020); Wong et al. (2004) | Liguoro et al. (2020); Yu and Chen (2020); Rajapakse and Dixit (2020) |
| 5 | 40 | 0.983 | 2016 | **neurological manifestation**; neurological complication; neurological involvement | Mehta et al. (2020); Mao et al. (2020a); Hamming et al. (2004) | Aghagoli et al. (2020); Munhoz et al. (2020); Chen et al. (2020d) |
| 6 | 37 | 0.999 | 2003 | **molecular epidemiology**; epidemic diarrhoea virus | Bosch et al. (2003) | Song and Park (2012) |

## 4. Summary and conclusions

As established by the current study, the collective response of the scientific community to the outbreak of SARS-CoV-2 has been unique, swift and arguably unprecedented in the history of scientific publishing in terms of its magnitude. While the analyses demonstrated that major outbreaks of coronavirus diseases do generate spikes of scholarly publications on their respective topics, the scale of the reaction from the scientific community to the most recent coronavirus outbreak has proven rather incomparable to the previous instances. An influx of scholarly publications following the recent outbreak has more than tripled the size of the coronavirus literature within a time span of nearly eight months, for a literature that has been in the making for more than 50 years and has already experienced two other bursts of activities as a result of the previous viral outbreaks. In fact, in many ways, the scientific literature of coronaviruses has proven to be a unique anomaly, featuring effects that have barely been observed in relation to other domains of science. The literature of most topics of science accumulate and develop gradually, often at an increasing rate. There are not many areas of science for which more than two thirds of the literature has been developed in less than a single year. In fact, the sheer size of this literature, per se, has now become rather enormous compared to the literature of many other scientific fields. As another feature, one can point to the creation and existence of a multitude of sleeping beauties in this literature. Sleeping beauties are generally rare in science, yet this field has created several of them already. Another feature is the retraction of articles. Retracted articles are generally rare and when they happen, it is often several years after the paper has been published. And yet, this field has created seven of them within less than eight months.

According to our analyses, the 2020 sector of this literature displays an unusually poor bibliographic connectivity level compared to the rest of the body of the literature. This could, in fact, be viewed as an inevitable feature when several thousands of articles are deployed in a literature almost concurrently within a time span of eight months. With the rapid developments that are currently occurring in the structure of this literature, the role of research synthesis has become more critical than ever. With more than 900 review and meta analyses already published on this topic since January 2020, it appears that the role of information and finding synthesis is, in fact, being increasingly recognised by the scientific community, which could, in time, compensate for the aforementioned effect.

Here, using scientometric methods, the structure and temporal development of the coronavirus literature, along with certain subsets of this literature (i.e., that of SARS-CoV-2, and that of reviews and meta analyses) were analysed at the level of individual articles. The composition of this vast literature (consisting of more than 44,000 published and officially indexed items) was analysed by identifying clusters/streams of highly connected or thematically similar research activities based on the number of references shared by individual studies or the frequency of individual studies being jointly cited by third articles. Individual subsets of studies were identified that have played a centrality role between the bodies of literature published prior to and after 2020. The oldest and newest streams of research activities, those that have largely terminated and those that are actively ongoing were identified through a document co-citation analysis. Based on the metrics of total citations, and local citations as well as citations bursts, most influential articles of this literature within each stream of activities were determined. Further co-citation analysis on the subset of reviews and meta-analytic studies of this literature identified the body of research synthesis activity in this domain that has been undertaken since January 2020 in contrast with the preceding years. It also identified review and meta studies with the highest relative coverage of the literature as well as individual articles that have most frequently been cited by reviews and meta analyses in various clusters of this literature.

An analysis on the subset of SARS-CoV-2 literature, also identified individual articles outside this literature; i.e., published prior to 2020, many of which having remained relatively unnoticed until now when currently proving instrumental in informing the emerging 2020 literature of SARS-CoV-2. The determination of such articles, some proving to be sleeping beauties of this literature, shows how research undertaken prior to a health emergency could suddenly serve scientists in the wake of unanticipated health crises. This could be a promising sign that, with the abundance of studies being currently undertaken during the SARS-CoV-2 crisis, the human community will be better prepared for responding to future disease outbreaks of respiratory nature. While the nature and chemistry of different diseases differ from one another and this often requires their own specific knowledge base, the integration established between health scientists and social scientists during the SARS-CoV-2 pandemic could be one lasting effect and a timeless impact of the current scientific efforts (Betsch, 2020; Van Bavel et al., 2020). This has been aptly pointed out by Leslie et al. (2020) who delineate how, facing disease epidemics of such magnitude "health systems might leverage social science to improve their preparations and communications".

It is hoped that these large-scale analyses of the coronavirus literature in its full scope have been able to provide useful insights into the nature and unique features of this literature, as well as its structural composition, temporal evolution and influential entities. In the face of the abundance of the studies already accumulated in this field, this could assist reviewers and authors to take better notice of influential work in their particular subdomains of interest within this field. The analyses could also be helpful in identifying areas where synthesis of research in the form of reviews or meta analyses could be needed or be more of the interest of researchers of this field. The cluster information presented alongside the analysis of each dataset could particularly assist scoping and narrative reviews on specific topics of this field.

## Acknowledgments

This research was funded by Australian Research Council grant DE210101175.

**Appendix**— Documents with longest citation bursts in the coronavirus literature

| References | Year | Strength | Begin | End | 1990 - 2020 |
|---|---|---|---|---|---|
| BRIERLEY I, 1989, CELL, V57, P537, DOI | 1989 | 42.6441 | 1990 | 2009 | |
| YEAGER CL, 1992, NATURE, V357, P420, DOI | 1992 | 77.1635 | 1993 | 2012 | |
| DELMAS B, 1992, NATURE, V357, P417, DOI | 1992 | 73.4639 | 1993 | 2012 | |
| Herrewegh AAPM, 1998, J VIROL, V72, P4508, DOI | 1998 | 65.9911 | 1999 | 2018 | |
| BOURSNELL MEG, 1987, J GEN VIROL, V68, P57, DOI | 1987 | 81.4751 | 1990 | 2007 | |
| Lai MMC, 1997, ADV VIRUS RES, V48, P1, DOI | 1997 | 171.2779 | 1999 | 2016 | |
| Gosert R, 2002, J VIROL, V76, P3697, DOI | 2002 | 62.7227 | 2003 | 2020 | |
| DEGROOT RJ, 1987, J MOL BIOL, V196, P963, DOI | 1987 | 37.6269 | 1990 | 2006 | |
| CAVANAGH D, 1988, VIRUS RES, V11, P141, DOI | 1988 | 25.4347 | 1990 | 2006 | |
| GORBALENYA AE, 1991, FEBS LETT, V288, P201, DOI | 1991 | 10.312 | 1993 | 2009 | |
| Bosch BJ, 2003, J VIROL, V77, P8801, DOI | 2003 | 94.2841 | 2004 | 2020 | |
| SAWICKI SG, 1990, J VIROL, V64, P1050, DOI | 1990 | 89.3478 | 1990 | 2005 | |
| SETHNA PB, 1989, P NATL ACAD SCI USA, V86, P5626, DOI | 1989 | 75.3769 | 1990 | 2005 | |
| FRANA MF, 1985, J VIROL, V56, P912, DOI | 1985 | 42.9792 | 1990 | 2005 | |
| VENNEMA H, 1990, J VIROL, V64, P339, DOI | 1990 | 29.8556 | 1991 | 2006 | |
| SNIJDER EJ, 1990, NUCLEIC ACIDS RES, V18, P4535, DOI | 1990 | 12.793 | 1991 | 2006 | |
| OLSEN CW, 1992, J VIROL, V66, P956, DOI | 1992 | 13.588 | 1993 | 2008 | |
| Vennema H, 1996, EMBO J, V15, P2020, DOI | 1996 | 79.8668 | 1997 | 2012 | |
| KUBO H, 1994, J VIROL, V68, P5403, DOI | 1994 | 50.4592 | 1997 | 2012 | |
| Cavanagh D, 1997, ARCH VIROL, V142, P629 | 1997 | 73.1325 | 1998 | 2013 | |
| Ziebuhr J, 2000, J GEN VIROL, V81, P853, DOI | 2000 | 94.908 | 2001 | 2016 | |
| SPAAN W, 1988, J GEN VIROL, V69, P2939, DOI | 1988 | 140.723 | 1990 | 2004 | |
| STURMAN LS, 1985, J VIROL, V56, P904, DOI | 1985 | 59.214 | 1990 | 2004 | |
| LUYTJES W, 1987, VIROLOGY, V161, P479, DOI | 1987 | 47.9584 | 1990 | 2004 | |
| JIMENEZ G, 1986, J VIROL, V60, P131 | 1986 | 29.0068 | 1990 | 2004 | |
| TOOZE J, 1984, EUR J CELL BIOL, V33, P281 | 1984 | 24.862 | 1990 | 2004 | |
| CAVANAGH D, 1986, J GEN VIROL, V67, P1435, DOI | 1986 | 24.2699 | 1990 | 2004 | |
| DELMAS B, 1990, J VIROL, V64, P5367, DOI | 1990 | 26.3812 | 1991 | 2005 | |
| DVEKSLER GS, 1991, J VIROL, V65, P6881, DOI | 1991 | 76.3331 | 1992 | 2006 | |
| WILLIAMS RK, 1991, P NATL ACAD SCI USA, V88, P5533, DOI | 1991 | 65.475 | 1992 | 2006 | |
| BARIC RS, 1988, J VIROL, V62, P4280, DOI | 1988 | 27.9393 | 1992 | 2006 | |
| SCHULTZE B, 1991, J VIROL, V65, P6232, DOI | 1991 | 17.1915 | 1992 | 2006 | |
| LIU DX, 1991, VIROLOGY, V184, P531, DOI | 1991 | 10.2067 | 1992 | 2006 | |
| van den Hoogen BG, 2001, NAT MED, V7, P719, DOI | 2001 | 97.0739 | 2003 | 2017 | |
| Cavanagh D, 2003, AVIAN PATHOL, V32, P567, DOI | 2003 | 41.9299 | 2004 | 2018 | |
| Li WD, 2005, SCIENCE, V310, P676, DOI | 2005 | 216.7862 | 2006 | 2020 | |
| Lau SKP, 2005, P NATL ACAD SCI USA, V102, P14040, DOI | 2005 | 206.1255 | 2006 | 2020 | |
| Weiss SR, 2005, MICROBIOL MOL BIOL R, V69, P635, DOI | 2005 | 76.3643 | 2006 | 2020 | |
| Simmons G, 2005, P NATL ACAD SCI USA, V102, P11876, DOI | 2005 | 66.5406 | 2006 | 2020 | |
| Brian DA, 2005, CURR TOP MICROBIOL, V287, P1, DOI | 2005 | 57.9637 | 2006 | 2020 | |
| PARKER SE, 1989, VIROLOGY, V173, P664, DOI | 1989 | 63.1798 | 1990 | 2003 | |
| LUYTJES W, 1988, VIROLOGY, V166, P415, DOI | 1988 | 53.8088 | 1990 | 2003 | |
| PENSAERT M, 1986, VET QUART, V8, P257, DOI | 1986 | 44.4062 | 1990 | 2003 | |
| DEGROOT RJ, 1989, VIROLOGY, V171, P493, DOI | 1989 | 11.9021 | 1991 | 2004 | |
| STOHLMAN SA, 1988, J VIROL, V62, P4288, DOI | 1988 | 23.7492 | 1992 | 2005 | |
| CHARLEY B, 1988, J VIROL, V62, P8, DOI | 1988 | 20.3557 | 1992 | 2005 | |
| YOKOMORI K, 1991, VIROLOGY, V183, P647, DOI | 1991 | 18.0936 | 1992 | 2005 | |
| JOHNSTON SL, 1995, BRIT MED J, V310, P1225, DOI | 1995 | 5.9402 | 1998 | 2011 | |
| Foley JE, 1997, J AM VET MED ASSOC, V210, P1307 | 1997 | 12.7448 | 1999 | 2012 | |
| Narayanan K, 2000, J VIROL, V74, P8127, DOI | 2000 | 31.7241 | 2001 | 2014 | |
| CARVAJAL A, 1995, J VET DIAGN INVEST, V7, P60, DOI | 1995 | 4.8502 | 2002 | 2015 | |
| Woo PCY, 2005, J VIROL, V79, P884, DOI | 2005 | 220.855 | 2005 | 2018 | |
| Casais R, 2003, J VIROL, V77, P9084, DOI | 2003 | 36.9944 | 2005 | 2018 | |
| Masters PS, 2006, ADV VIRUS RES, V66, P193, DOI | 2006 | 108.9006 | 2007 | 2020 | |
| Tang XC, 2006, J VIROL, V80, P7481, DOI | 2006 | 62.6439 | 2007 | 2020 | |
| FLEMING JO, 1986, J VIROL, V58, P869, DOI | 1986 | 58.1709 | 1990 | 2002 | |
| BRIERLEY I, 1987, EMBO J, V6, P3779, DOI | 1987 | 38.4497 | 1990 | 2002 | |
| FLEMING JO, 1983, VIROLOGY, V131, P296, DOI | 1983 | 37.6185 | 1990 | 2002 | |
| BAKER SC, 1989, J VIROL, V63, P3693, DOI | 1989 | 29.9653 | 1990 | 2002 | |
| LAI MMC, 1990, ANNU REV MICROBIOL, V44, P303, DOI | 1990 | 119.524 | 1991 | 2003 | |